\documentclass[1p]{elsarticle}
\usepackage{amssymb}
\usepackage{amsmath}
\usepackage{booktabs} 
\usepackage{graphicx}
\usepackage{dcolumn}
\usepackage{bm}
\usepackage{float}
\usepackage{placeins}
\usepackage{subcaption}
\usepackage{booktabs}
\usepackage{threeparttable}

\usepackage{threeparttablex}
\usepackage{longtable}
\usepackage{etoolbox}
\appto\TPTnoteSettings{\footnotesize}

\usepackage{subcaption}
\newcommand{\f}[2]{\frac{#1}{#2}}
\newcommand{\eq}[1]{\begin{gather}\begin{aligned} #1 \end{aligned} \end{gather}}

\newcommand{\fig}[4]
{\begin{figure}[!htb]
\centering
\includegraphics*[width=#4\textwidth]{Figures/#1}
\caption{#2 \label{#3}}
\end{figure}}
\newcommand{\wfig}[4]
{\begin{figure*}[!htb]
\centering
\includegraphics[width=#4\textwidth]{Figures/#1}
\caption{#2 \label{#3}}
\end{figure*}}

\newcommand{\subfig}[6]{
\begin{figure}
  \begin{subfigure}[b]{0.45\columnwidth}
    \includegraphics[width=\linewidth]{Figures/#1}
    \caption{}
    \label{#2}
  \end{subfigure}
  \hfill 
  \begin{subfigure}[b]{0.45\columnwidth}
    \includegraphics[width=\linewidth]{Figures/#3}
    \caption{}
    \label{#4}
  \end{subfigure}
  \caption{#5}
  \label{#6}
\end{figure}
}
\newcommand{\mt}[4]{
\begin{table}[!htb]
\begin{center}
\begin{tabular}{#1}
#2 
\end{tabular}
\caption{#4\label{#3}}
\end{center}
\end{table}}

\begin{document}
\begin{frontmatter}

\title{Tapered helical undulator system \\
for high efficiency energy extraction \\ from a high brightness electron beam}

\author[inst1]{Y. Park}
\author[inst2]{R. Agustsson}
\author[inst4]{W. J. Berg}
\author[inst4]{J. Byrd}
\author[inst2]{T. J. Campese}
\author[inst1]{D. Dang}
\author[inst1]{P. Denham}
\author[inst4]{J. Dooling}
\author[inst1]{A. Fisher}
\author[inst2]{I. Gadjev}
\author[inst3]{C. Hall}
\author[inst1]{J. Isen}
\author[inst1]{J. Jin}
\author[inst4]{A. H. Lumpkin}
\author[inst2]{A. Murokh}
\author[inst4]{Y. Sun}
\author[inst5]{W. H. Tan}
\author[inst3]{S. Webb}
\author[inst4]{K. P. Wootton}
\author[inst4]{A. A. Zholents}
\author[inst1]{P. Musumeci}%

\affiliation[inst1]{organization={UCLA Department of Physics and astronomy},
            addressline={475 Portola Plaza}, 
            city={Los Angeles},
            state={CA},
            postcode={90095}, 
            country={U.S.A.}}
\affiliation[inst2]{ organization={RadiaBeam Technologies} ,
            addressline={1717 Stewart St,},
            city={Santa Monica},
            state={CA},
            postcode={90404},
            country={U.S.A.}}
\affiliation[inst3]{organization= {RadiaSoft},
            addressline={6640 Gunpark Dr Suite 200},
            city={ Boulder},
            state={CO},
            postcode={80301},
            country={U.S.A.}}
\affiliation[inst4]{organization={Argonne National Laboratory},
           addressline={9700 S Cass Ave},
           city={Lemont},
           state={IL},
           postcode={60439},
           country={U.S.A.}}%
\affiliation[inst5]{organization={Northern Illinois University},
            addressline={1425 W Lincoln Hwy},
            city={DeKalb},
            state={IL},
            postcode={60115},
            country={U.S.A.}}

\begin{abstract}
In this paper we discuss the design choices and construction strategy of the tapered undulator system designed for a high energy extraction efficiency experiment in the ultraviolet region of the electromagnetic spectrum planned for installation at the Argonne National Laboratory Linac Extension Area (LEA) beamline. The undulator is comprised of 4 sections pure permanent magnet Halbach array separated by short break sections, each one of them housing a focusing quadrupole doublet and a phase shifter. The quadrupoles use a novel hybrid design which allows one to vary the gradient and match the beam transversely. The undulator tapering profile is optimized to maximize the energy conversion efficiency from a 343~MeV 1~kA beam into coherent 257.5 nm radiation taking into account the longitudinal current profile generated by the linac.  
\end{abstract}

\end{frontmatter}

\section{\label{introduction}Introduction}

Modern particle accelerators are very efficient in converting and concentrating wall-plug electrical power into relativistic electron beams \cite{padamsee:SRF, bane:NCRF}. At the same time, converting the electron kinetic energy into coherent radiation in an efficient manner still remains an open challenge. Beam-matter interaction is inefficient, generates large amount of heat and the electromagnetic energy is degraded to incoherent radiation. Free-electron laser (FEL) schemes potentially have very low losses as the interaction between charged particles and radiation takes place in vacuum, but in their practical implementation have been so far still limited by saturation to efficiencies at short wavelength on the order of the FEL parameter $\rho$, which is typically well below 1$\%$ \cite{BPN}. Tapering the undulator (i.e. varying its parameters along the interaction) has been shown both in theory and experiment to improve the conversion efficiency of relativistic electron beam power into coherent short wavelength radiation in FELs \cite{KMR, orzechowski, bonifacio1989tapering, scharlemann1988selected, feldman1989high, hafizi1990efficiency, fawley:taperedsase, jiao2012modeling, emma:tapering,  makcurbis, schneidmiller:tapering}. This approach has tremendous potential to enable fundamental breakthroughs in various areas of science and industry \cite{fratalocchi, schwinger:XFEL, pagani:EUVlitho, hosler:EUVFEL, murokh:EUVlitho} owing to the fact that high efficiency FELs are essentially ideal light sources, capable of very high average and peak power (only limited by beam power) and tunable over the entire region of the electromagnetic spectrum by simply changing the e-beam energy. 


The physical concept that we pursue in the quest for very high efficiency FELs is the so-called Tapering Enhanced Stimulated Superradiant Amplification (TESSA in short) regime where high intensity seed and pre-bunched electron beams are used in combination with strongly tapered undulators to sustain high gradient deceleration over extended distances, resulting in the conversion of most of the energy from the beam into coherent radiation \cite{JDuris2015, gover:RMP}. In a TESSA system high gradient deceleration takes place at very high energy exchange rate (up to ~10 MV/m peak gradients), nearly one order of magnitude larger than in any known FEL \cite{orzechowski, hajima:high_efficiency}, in order to beat the onset of sideband instabilities which have been known for decades to set the limit on tapered FEL energy exchange \cite{Colsonsideband, davidson1987single, tsai2017sideband}. In a previous experiment in the TESSA regime \cite{NSudar2016}, efficiencies as high as 30 $\%$ (at 10 $\mu$m wavelength) have been demonstrated. This experiment was carried out using an intense external seed in a very low gain amplification regime. The presence of the strong background signal from the seed laser precluded a full characterization of the transverse and spectral profiles of the amplified radiation. 

In this paper we describe TESSA-266, a seeded FEL experiment aiming at fundamentally addressing the efficiency limitations in electron-based coherent radiation generation by exploiting recent progress in high brightness beam sources \cite{musumeci2018advances} and increased understanding of the strong coupling with the electromagnetic field in tapered undulator systems \cite{JDuris2015, emma2017high}. The goal of the experiment is to demonstrate record high single pass energy extraction efficiency in the strongly tapered seeded regime at Ultra-Violet (UV) wavelengths. The program is the result of a collaborative effort of UCLA, RadiaBeam, Argonne National Laboratory and RadiaSoft and has been developed based on the possibility to use the high brightness beam from the Advanced Photon Source (APS) linac, an electron accelerator with energy up to 500 MeV, and nominally operated around 350 MeV \cite{lewellen1999_apslinac}. Compared to previous tapered undulator experiments, the main novelties in the setup are the use of a strongly tapered helical undulator to provide continuous coupling with circularly polarized electromagnetic waves and a novel focusing scheme based on the implementation of a quadrupole doublet in the undulator break section to maintain very high beam density and maximize the radiation generation. 


In the paper we discuss the different components of the undulator system for the TESSA-266 experiment, starting with an overview of the parameter choices (Section II) which includes the results of the numerical simulations utilized in the design to quantify the limits associated with the incoming beam quality. We then delve into the technical details of the magnetic undulator (Section III) and the break-section design (Section IV). In order to keep as short as possible the total length of the system and meet the physics and engineering constraints for high efficiency amplification, we adopted innovative technical solutions which might find application in other FEL developments. In addition, this paper serves to the TESSA-266 collaboration as a repository of design information and to the general reader as a case study reviewing all the required steps in the design of a high efficiency FEL system. 


\section{\label{Overview} Overview of the experiment}
\begin{figure*}[ht]
\centering
\includegraphics[width=\textwidth]{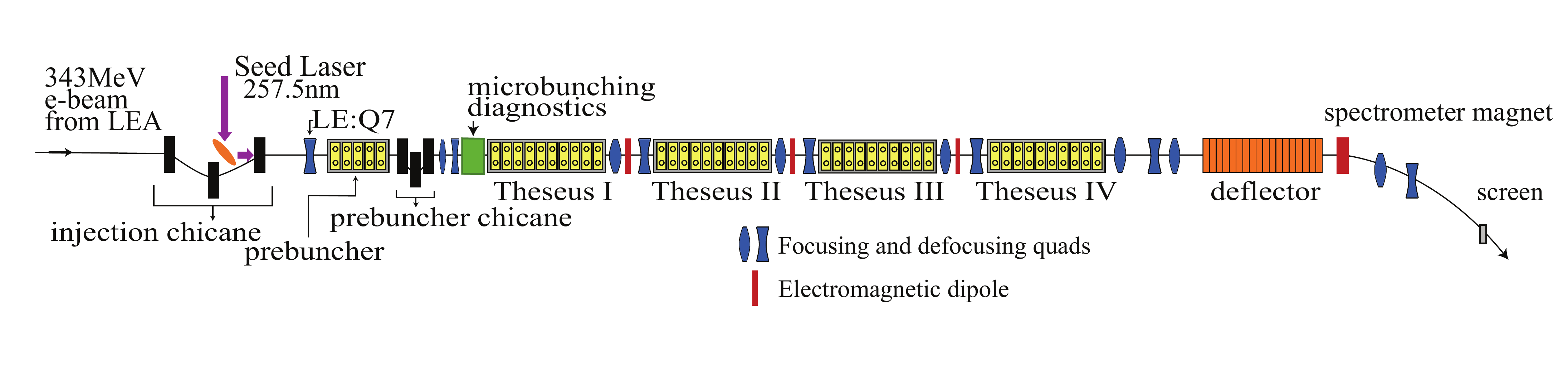}
\caption{A schematic diagram of TESSA-266 beamline. The beam moves from left to right. The injection chicane creates an offset in the electron beam trajectory to allow the on axis injection of the input seed laser. A modulator-chicane module is used to prebunch the beam at the seed wavelength. The energy is then extracted from the beam using four 1 m long undulator sections, separated by a quadrupole doublet to control the transverse beam size. The post-undulator diagnostics section includes matching quadrupoles, a horizontal deflecting cavity and a vertically dispersing dipole for longitudinal phase space measurements.}
\label{fig:expsetup}
\end{figure*}

For sake of initial design, we will assume a 343 MeV energy electron beam with 300 pC beam charge 1.0 kA peak current, 2 mm-mrad normalized transverse emittance and 300 keV RMS incoherent energy spread. In principle, these could be provided by operating the APS linac with the high gradient RF photocathode (PC) gun to inject high brightness beams in the LEA beamline. Operation of the APS ring would not be impacted by using a fast automated switching between the PC gun and the thermoionic sources required to fill the ring \cite{Shin2018}. The maximum repetition rate for beam delivery in the LEA beamline is 10 Hz. The design parameters for the experiment are summarized in Table \ref{tab::LEAbeam} and a schematic of the TESSA-266 beamline is shown in Fig. \ref{fig:expsetup}.


\begin{table}[!hbt]
	\centering
	\begin{tabular}{l|c c }
	\toprule
Parameter & Unit & Value \\
\hline
Beam Energy & MeV & 343  \\
Bunch Charge & pC& 300  \\
Peak Current & A &  1000  \\
Bunch Length, FWHM & ps & 0.3  \\
Normalized Emittance & mm-mrad & 2 \\
Uncorr. energy spread, FWHM & keV & 300 \\
Rel. energy variation along bunch & \% & $<$0.5\\
Linac rep rate & Hz & 10 \\ 
\hline
Seed laser wavelength & nm & 257.5 \\
Seed laser peak power & GW & 1 \\
Seed laser pulse energy & mJ & 0.5 \\
\bottomrule
\end{tabular}
	\caption{TESSA-266 Design Parameters \label{tab::LEAbeam}}
\end{table}

In order to facilitate the energy extraction from the beam, the TESSA concept relies on an intense seed pulse which in this case is provided by an external laser system. The wavelength selection for the experiment is in fact driven by the desire to demonstrate the high gain TESSA regime at the shortest possible wavelength where intense seed laser pulses can be easily available. For shorter wavelengths, more complex configurations like higher harmonic generation or a TESSA-based oscillator (TESSO) need to be employed \cite{TESSO}. In our baseline design, a Yb-based master-oscillator power-amplifier system provides 10 Hz 5 mJ 0.5 ps 1030 nm pulses which are synchronized with the e-beam arrival time. Two non-linear frequency doubling crystals 0.9 mm thick are then used to obtain 1 mJ / 0.5 ps 257.5 nm pulses. The laser is then deflected to be co-propagating with the e-beam in the TESSA-266 beamline using a small on-axis mirror located in the vacuum chamber of the injection chicane as shown in Fig. \ref{fig:expsetup}. We assume that up to 1 GW peak power can be available in the seed pulse after taking into account transport mirror reflectivity and losses on the vacuum window.

A prebunching scheme is used to prepare the e-beam to be loaded in the ponderomotive bucket of the tapered undulator. The laser pulse overlaps temporally and spatially with the electron beam in a 0.324~m long helical undulator (period length 32~mm) which serves to modulate the electron beam energy. The resulting energy modulation amplitude (up to 2.8 MeV peak-to-peak), is then converted into density modulation by a tunable R56 (10-20 $\mu$m) chicane prior injecting the beam at the right phase with respect to the seed pulse into the TESSA-266 gap-tapered undulator. A microbunching diagnostic chamber \cite{lumpkin2001first, lumpkin2002evidence, lumpkin2003transverse} is to be installed in the section between the prebuncher and the main undulator to optimize the bunching factor. The microbunching diagnostics is based on the generation of coherent optical transition radiation (COTR) which would be enhanced by almost seven orders of magnitude over incoherent OTR with only a 10$\%$ microbunching fraction. By using beam splitters with optical paths to multiple UV sensitive cameras to record the intensity of the near and far field of the radiation, it will be possible to assess the microbunched electron beam transverse size, angular divergence, spectrum, microbunching fraction. In addition, overlap of the seed laser and the electron beam in the modulator could be tuned (scanning the relative time of arrival over multiple shots) based on the observed distributions \cite{lumpkin:FEL,lumpkin:IBIC}. The design value for the input bunching factor is 0.5, but even larger bunching factors could be obtained using a higher laser intensity in the modulator. Diffraction effects for the seed laser pulse suggest a compromise on the laser waist position between having strong intensity at the buncher or at the entrance of the tapered undulator. As discussed below, a numerical study indicates that the optimal focusing conditions for the laser are a waist position plane $z_w$ at the entrance of the tapered undulator, and a Rayleigh range of 1.45 m. 

The THESEUS (Tapered HElical SEgmented Undulator System) undulator sections are helical undulators with four arrays of magnets in Halbach structure. The helical geometry increases by more than a factor of two the FEL coupling compared to the corresponding planar case \cite{duris2012}. The undulator is formed by four 28 period long (with period length $\lambda_u = 32~$mm) sections which are interspaced by short break sections (225 mm long) equipped with doublet quadrupole focusing, vacuum pumps and beam and radiation diagnostics. The total length of the undulator system (prebuncher entrance to 4th undulator exit, not including the 2.15 m long mini-chicane for laser injection) is 6.26 m. 

After the undulator a diagnostic section to measure the longitudinal phase space of the TESSA-266 output beam has been foreseen. The main components of this section are a deflecting cavity and a dipole spectrometer. Temporally streaking the beam will be important to study the time-dependent dynamics of the radiation amplification, while the spectrometer needs to be able to both resolve the input beam energy spread, as well as capture the entire energy spectrum after deceleration. A series of quadrupoles is added to control the evolution of the transverse beam parameters along the beamline and maximize the energy and temporal resolution at the final screen. We assume a deflecting voltage of $<$ 10 MV/m and optimize the beta functions at the screen to be able to record the final image of the longitudinal phase space with 30 fs temporal resolution and 2e-4 energy resolution. Since the total beam energy spectrum is very wide after the undulator, the acceptance of the energy spectrometer system will be $>$ 15 \%.

\subsection{\label{Simulations} Tapering design and simulations}

There are several parameters in the design of the TESSA-266 experiment to be optimized in order to maximize the conversion efficiency. Using time-independent single frequency \textsc{genesis} simulations -- i.e. assuming infinitely long electron and laser pulses --, a numerical optimization maximizes the bunching factor at the undulator entrance as a function of seed laser Rayleigh length and waist position as well as the prebuncher chicane R56 setting. The resonant phase and optimum tapering design along the undulators are then determined using the Genesis-Informed-Tapering (GIT) algorithm \cite{JDuris2015} in order to maximize the radiation power at the exit of the system. Effectively the algorithm maximizes the number of particles trapped in the ponderomotive bucket and the deceleration gradient along the interaction. The phase shifts in between the undulator sections (to compensate diffraction and dephasing effects) are also separately optimized. Fig. \ref{fig::tindep} shows that the maximum power obtained in these simulation studies is 39.5 GW which translates to 11 \% peak conversion efficiency, consistent with the average change in the beam energy (right panel). The resonant energy along the interaction decreases by almost 20 $\%$ as can be seen looking at the undulator normalized parameter $K$, but not all particles participate to the interaction and the bunching factor remains around 0.6 for the entire undulator. 

\wfig{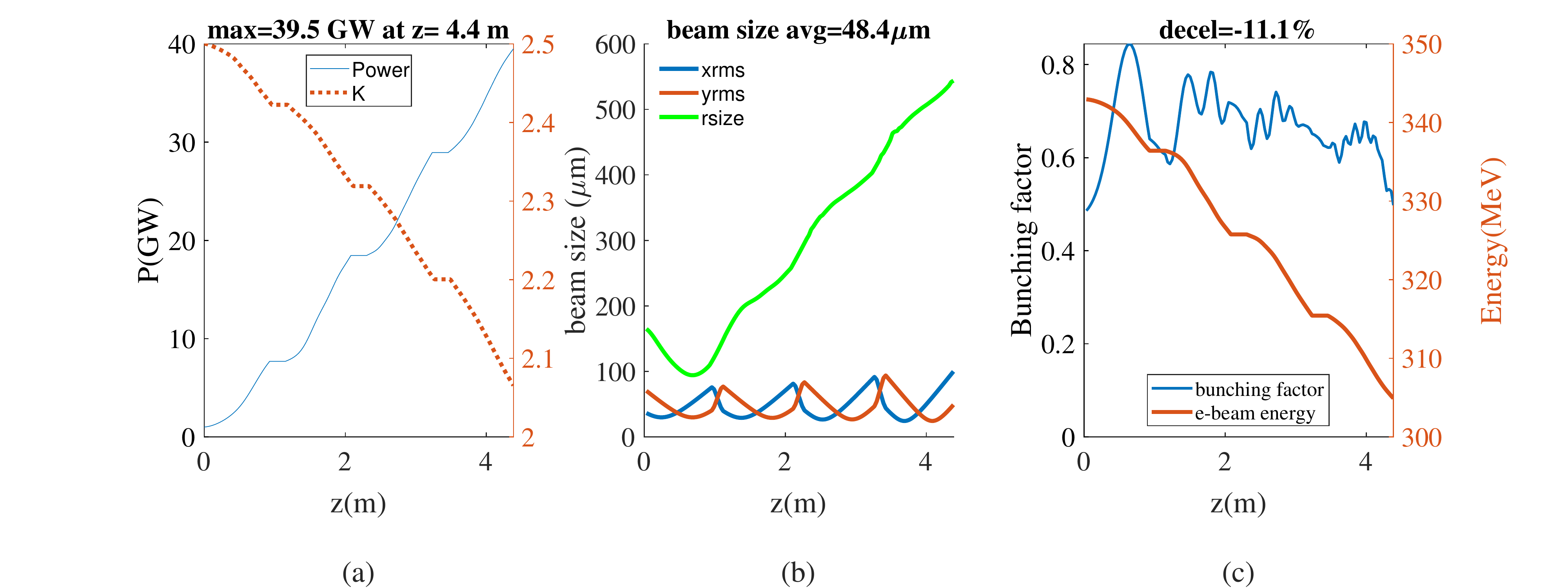}{Results of the numerical optimization of the tapering using the Genesis-Informed-Tapering algorithm. (a) Radiation power and normalized undulator parameter $K$, (b) radiation and e-beam horizontal and vertical spot sizes, (c) bunching factor and average beam energy plotted along the undulator.}{fig::tindep}{1}

The transverse lattice of the TESSA-266 beamline was optimized to maximize the extraction efficiency in the strongly tapered undulators. Due to the inverse-square dependence of efficiency on transverse beam size \cite{emma2017high}, solutions on how to minimize the average beam size along the interaction were sought \cite{YPark2018}. Assuming a 2 mm-mrad emittance, with no added quadrupoles, solely relying on the undulator focusing the average rms size in the undulator can be calculated to be $\langle \sigma_x \rangle=80\mu$m. The standard solution of introducing alternating gradient focusing and defocusing quadrupoles in each drift sections yields a $\langle \sigma_x \rangle=68\mu$m. It was then found that using a quadrupole doublet in the break section would make the smallest transverse average rms beam size of 46$~\mu$m (Fig. \ref{fig::beamsize}). The different focusing lattice solutions were compared using the time-independent tapering optimization algorithm, and the solution with a quadrupole doublet in the break section was found to yield the smallest average transverse beam size in the undulators and the highest output power and was decided as the design choice. 

The beam from the linac can be matched to this strong focusing channel using two electromagnetic quadrupoles in between the prebuncher and the first section. A list of all the magnetic elements in the TESSA-266 beamline is reported in Table \ref{tab:lattice}. Besides the undulator and the quadrupole doublet in the break sections, this list also includes the dipoles in the injection and modulation chicanes and matching quadrupoles before and after the prebuncher. 
\fig{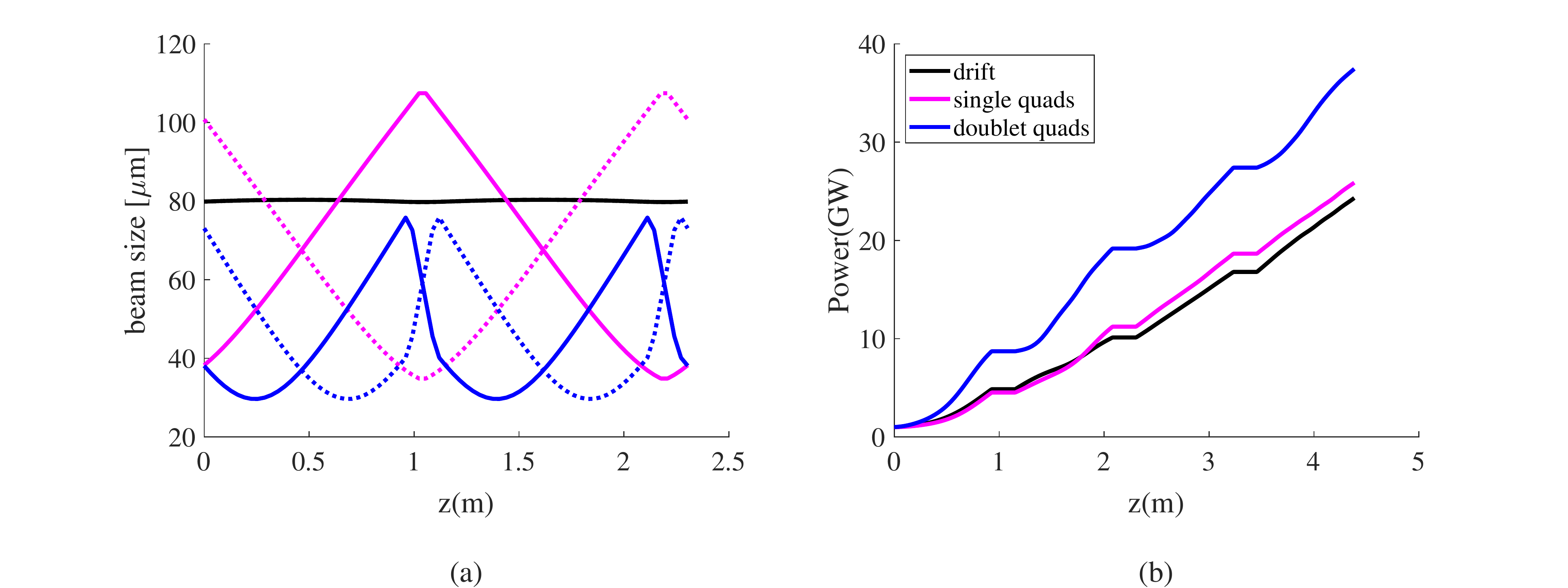}{(a) Horizontal (solid) and vertical (dotted) transverse beam sizes along the undulator for the various lattices considered in the design phase assuming a matched 343 MeV electron beam with initial emittance 2 mm-mrad and 29$\lambda_u$-long undulators and 7$\lambda_u$ break sections. Compared to other schemes such as relying solely on the undulator natural focusing without any quadrupoles (black), and single alternating quads in the drift (magenta), the quadrupole doublet solution (blue) yields the smallest average size in the undulator and the largest output power. (b) Output power of drift (black), single alternating quads (magenta), quadrupole doublet (blue)  lattices with tapering optimized in time-independent scheme.}{fig::beamsize}{1}

    \begin{longtable}{|c|c|c|c|}
    \pagebreak
    
\caption{List of TESSA-266 beamline elements including magnetic lengths, magnetic strengths and distances from the last reference quadrupole (LE:Q7) on the LEA beamline. The third column specifies the magnetic field for the dipoles and the integrated gradient for the quadrupoles. The physical length of the undulators is 25 mm longer than the one specified due to the end plates\label{tab:lattice}}\\
\hline
\multicolumn{4}{c}{TESSA-266 Beamline} \\
\hline
\hline
Element Name & Length & B[T] & Center from LE:Q7 \\
\hline
\hline
\multicolumn{4}{|c|}{Injection and Prebuncher} \\
\hline
Injection Chicane Dipole 1 & 0.146 & 0.09 & 1.78\\
Injection Chicane Dipole 2 & 0.146 & 0.18 & 2.80 \\
Injection Chicane Dipole 3 & 0.146 & 0.09 & 3.82 \\
\hline
LE:Q8 & 0.153 & 0.189 & 4.40\\
\hline 
\hline
Prebuncher & 0.324 & & 5.20 \\
\hline
Prebuncher Chicane Dipole 1 & 0.146 & 0.08 & 5.60 \\
Prebuncher Chicane Dipole 2 & 0.146 & 0.16 & 5.75 \\
Prebuncher Chicane Dipole 3 & 0.146 & 0.08 & 5.90 \\
\hline
TE:Q1 & 0.153 & 2.8  & 6.08 \\
TE:Q2 & 0.153 & -2.8  & 6.30 \\
Microbunching Diagnostics & 0.340& &\\
\hline
\hline
\multicolumn{4}{|c|}{THESEUS Undulators} \\
\hline
THESEUS I& 0.964 && 7.19 \\
\hline
TE:Q3 & 0.0307 &5.62 & 7.72\\
TE:D1 & 0.0263 & 0.25 & 7.76 \\
Diagnostics Cross:YAG2 &0.0254 & &7.80 \\
TE:Q4 & 0.0307 & 5.62 & 7.82 \\
\hline
THESEUS II & 0.964 && 8.35 \\
\hline
TE:Q5 & 0.0307  & 5.62 & 8.88\\
TE:D2 & 0.0263 & 0.25 & 8.92 \\
Diagnostics Cross:YAG3 &0.0254 & & 8.95 \\
TE:Q6 & 0.0307  & 5.62  & 8.98 \\
\hline
THESEUS III & 0.964 && 9.51 \\
\hline
TE:Q7 & 0.0307  & 5.62  & 10.04\\
TE:D3 & 0.0263 & 0.25 & 10.08\\
Diagnostics Cross:YAG4 & 0.0254& & 10.11\\
TE:Q8& 0.0307  & 5.62 & 10.14 \\
\hline
THESEUS IV & 0.964 && 10.67 \\
\hline
\hline
\multicolumn{4}{|c|}{Post-Undulator Diagnostics} \\
\hline
TT:Q1 & 0.104&-2.85& 11.4 \\
TT:Q2 &0.104&2.26& 12.2 \\
TT:Q3 &0.104&-1.76& 12.5 \\
Deflecting cavity &1.2&& 13.3 \\
Spectrometer magnet &0.25&0.4& 14.2 \\
TT:Q4 &0.104&-1.08& 14.5 \\
TT:Q5 &0.104&1.34& 15.0 \\
Diagnostics Cross:YAG5 &&& 16.7 \\
DUMP &&& 18.9\\
\hline
\end{longtable}


\subsection{\label{lps} Time-dependent simulations and longitudinal phase space matching}

While optimizing the tapering for realistic beam distributions has the potential to improve the results by taking into account the actual energy and temporal profiles of the electrons, time-dependent simulations are computationally intensive and not suited to large parameter scans. For this reason, in order to evaluate finite bunch length effects and provide guidance to the optimization of the beam dynamics through the linac, we employed the tapering solution obtained from the time-independent simulations. Due to the slippage phenomenon, it is found that flat current distributions (Figure \ref{fig::tdeprec}) yield a significantly higher FEL efficiency compared to Gaussian distributions. In Fig. \ref{fig::tdep} we show the output for two electron bunches having identical rms peak current and charge, but different current profiles. It is immediate to see how the radiation pulse overlaps better and for longer with the e-beam current distribution for the flatter current distribution. The energy conversion efficiency can be obtained by the ratio of the total energy in pulse (integral of the power profile along the bunch coordinate) to the e-beam energy (300 pC 343 MeV corresponding to 100 mJ). The flat current profiles shows a significantly larger conversion efficiency 8 $\%$ with respect to the 5 $\%$ obtained using a gaussian pulse. In the limit of very long flat pulse, the 10 $\%$ conversion efficiency limit of the steady state would be recovered. 
Including in the simulations resistive wall wake effects produced only a small change in the power output. For example, considering a gaussian profile bunch in a stainless steel pipe, time-dependent \textsc{genesis} simulations show a minimal reduction in efficiency from 5.23\% to 5.14\%. Surface roughness effects are harder to estimate as they depend on the vacuum pipe quality. Assuming similar order of magnitude effect, we don’t expect the experiment results to be significantly affected by wakefields.

\subfig{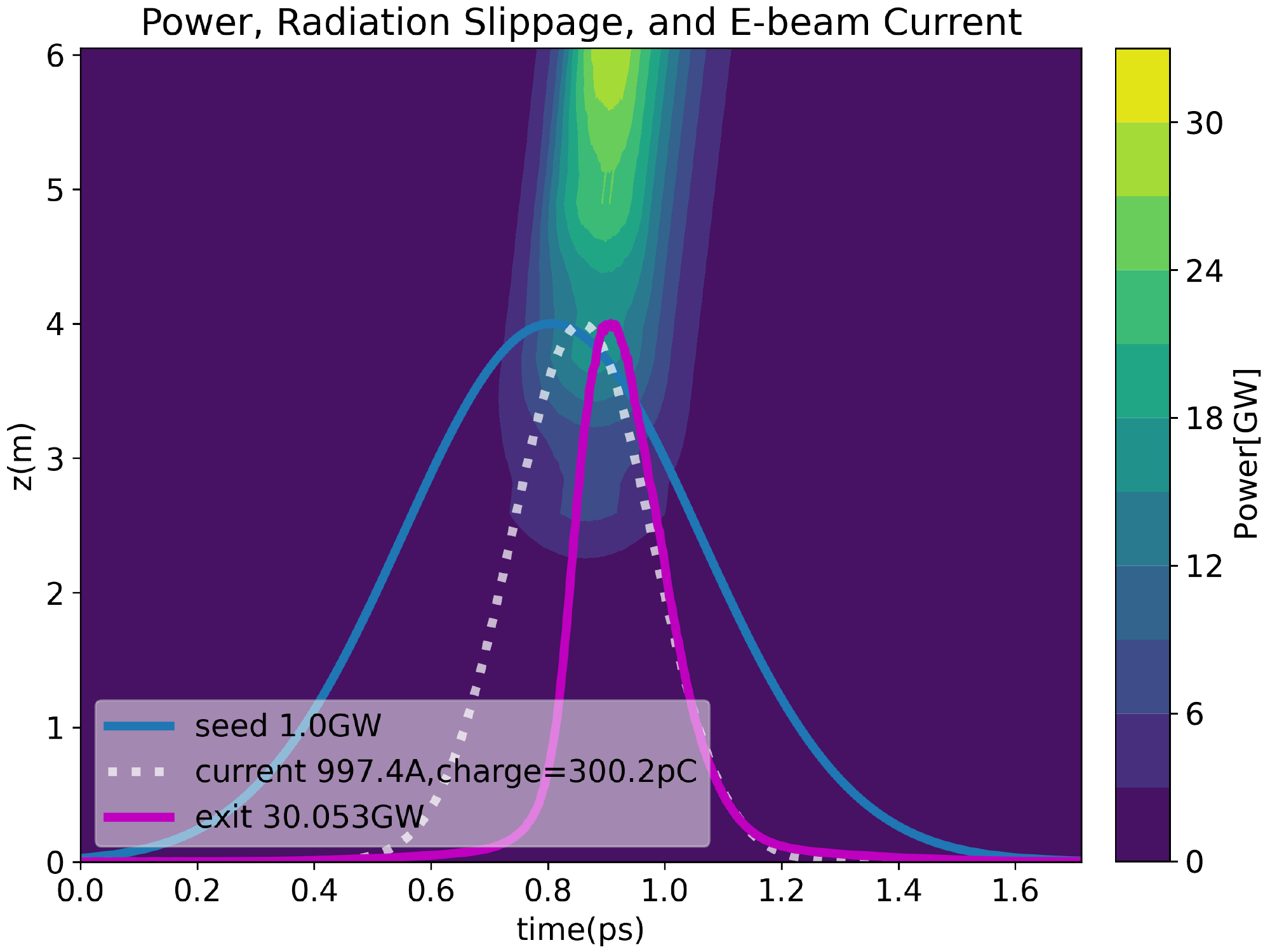}{fig::tdep}{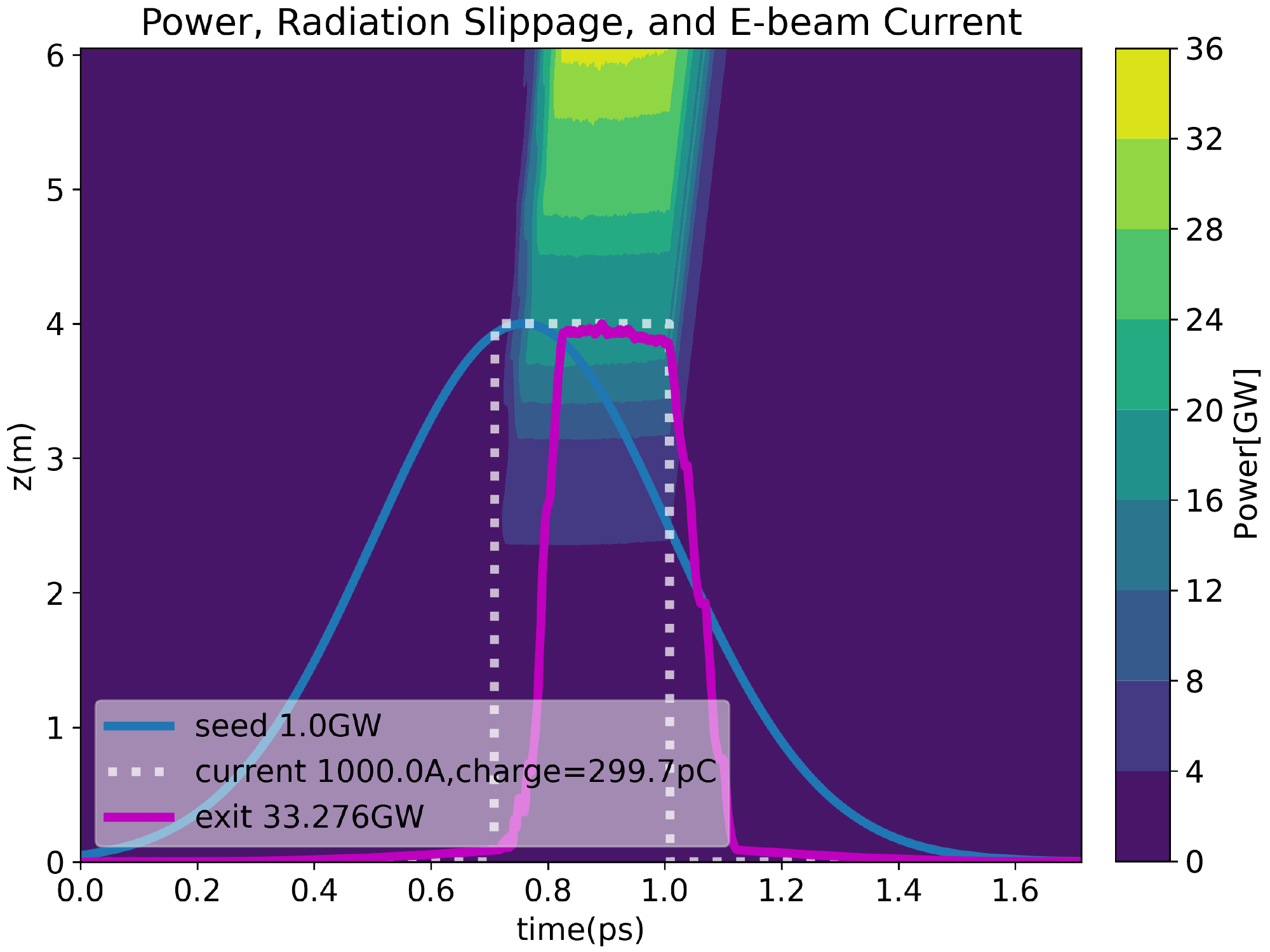}{fig::tdeprec}{ Waterfall plots of radiation power output along the undulator for time-dependent \textsc{genesis} for an input beam having Gaussian (left) and rectangular (right) temporal profiles with 300~pC charge, 1~kA peak current showing 5\% and 8 \% efficiency respectively. The input current (dotted yellow) and input seed power (blue) profiles are also shown. The final radiation profiles are shown in magenta solid lines.}{fig::tdepgen}
A careful management of the injected longitudinal phase space will be required in order to maximize the efficiency for TESSA-266. The way this is accomplished would entirely depend on the available linac system, but here we briefly discuss the case where the experiment is installed at the end of the APS Linac \cite{milton2001exponential} and driven by the high brightness photocathode RF gun (PC) recently been installed as a secondary APS injector \cite{sun2018high, Shin2018}. 

\begin{figure*}[ht]
\includegraphics[width=\textwidth]{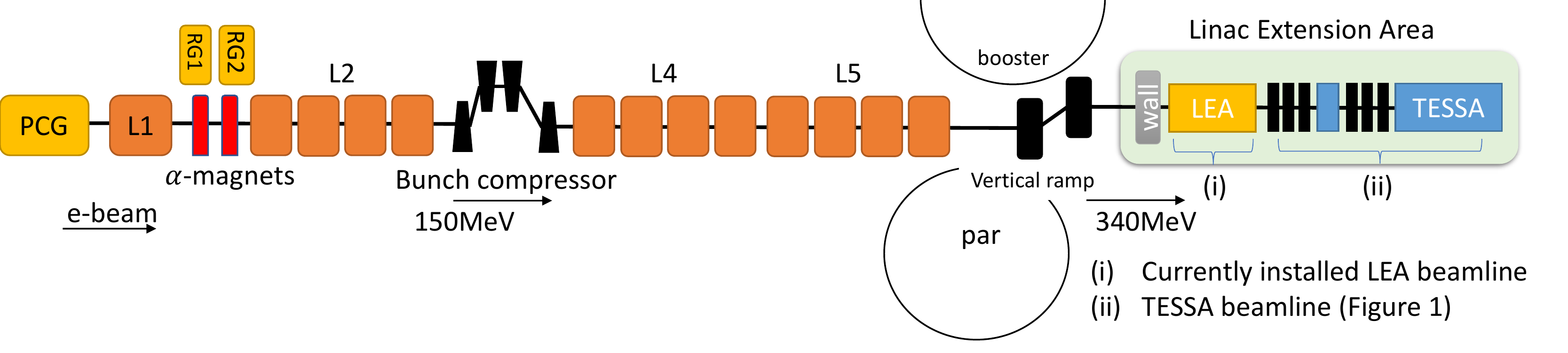}
\caption{Schematic of APS linac to TESSA-266 beamline for longitudinal phase space studies. Electron beams injected from PC Gun are accelerated through four main linacs L1, L2, L4, and L5. Electron beam accelerated to 150MeV by L1 and L2 become compressed by a magnetic chicane. L4 and L5 can accelerate the beam further to 343MeV. The beam passes through the vertical ramp magnets and enter the linac extension area at the end of which the TESSA-266 experiment can be installed.}
\label{fig:LEAbeamline}
\end{figure*}

Taking as a reference Fig. \ref{fig:LEAbeamline}, electrons from the PC gun accelerate through four linac structures L1, L2, L4, and L5\cite{Berg:LEA}. A bunch compressor is located just after the L2 linac section \cite{BorlandBC2000}. Through the L4 and L5 linac sections, the electrons are accelerated to the optimal energy for TESSA-266 experiment ($\gamma=671$). After the linear accelerator, the beam is deflected in a vertical ramp (R56=0.129mm, T566=30mm, U5666=281.8mm) which has a small effect on the final longitudinal phase space, before being sent thru the LEA beamline to the injection chicane (R56=-0.149mm, T566=0.224mm, U5666=-0.147mm) to create a 10~mm offset to enable the injection of the seed laser. A shortcoming of the current linac configuration is the lack of a high frequency RF linearizer that would greatly help in controlling the longitudinal phase space after compression. 

In order to better understand the challenge, we can model the energy-position correlation upstream of the bunch compressor--i.e. at the end of PC gun and linacs L1 and L2-- using the following expression
\eq{
\gamma_f(z) &= A_{\text{gun}}\sin(kz+\phi_g) ...\\
& +A_1\sin(kz +\phi_1) + A_2\sin(kz+\phi_2)
\label{eq::beamlineEq}
}
where $A_{\text{gun}}$, $A_1$, $A_2$ and $\phi_g$, $\phi_1$, and $\phi_2$ are the maximum energy gains and the phases of PC gun, L1, and L2 respectively and $k$ is wave vector of S-band structures. 

Assuming a short bunch, we can use a Taylor series expansion of Equation \ref{eq::beamlineEq}, and express the relative energy of each particle $p_i$ just upstream of the bunch compressor in a polynomial form. Keeping up to third order coefficients we have:
\eq{
p_i = d_1 z_i + d_2 (z_i)^2 + d_3 (z_i)^3 + p_{0,i}
}
A closer look to Eq. \ref{eq::beamlineEq} will show that regardless of the beam chirp, for all accelerating phases (between 0 and $\pi$) the energy position correlation is characterized by a negative second order curvature at the entrance at the bunch compressor. After compression, this term is responsible for a skewed, narrow current spike shown in Fig. \ref{fig::lps}a which would severely degrade the FEL interaction due to slippage even though the peak current is 1 ~kA.

In principle, it is possible to correct for the curvature if a non linear compression can be applied. In fact, the final longitudinal coordinates after the bunch compressor can be written as:
\eq{
z_{f_i} = R_{56}(p_i) + T_{566}(p_i)^2 + U_{5666}(p_i)^3
}
Combining the two last equations shows that the second order non linear coefficient would be cancelled if the following condition were satisfied: 
\eq{
-\f{T_{566}}{R_{56}} d_1^2 = d_2
}
Unfortunately in a standard magnetic chicane bunch compressor $R_{56}$ and $T_{566}$ are opposite in sign and the only way to satisfy this condition is to introduce non linear elements in the chicane (changing the sign of $T_{566}$ or adjust the curvature coefficient $d_2$ to be positive.

Both options have been evaluated using the start-to-end model developed starting with \textsc{astra} simulation up to L1 exit \cite{astra}, and then \textsc{elegant} to track the particles to TESSA-266 entrance \cite{borland_elegant}. We first considered adding sextupoles to control $T_{566}$ in the bunch compressor. Two sextupoles would be sufficient to fully linearize the longitudinal dynamics, but in order to avoid severe transverse emittance growth they should be placed at location with a $\pi$ phase advance between them \cite{england_sextupole, hall_sextupole}. However, due to the skewed design of the bunch compressor which was to minimize the CSR effects \cite{BorlandBC2000} along with the restrictions associated to satisfying the interleaving lattice matching \cite{Shin2018}, it proved too difficult to get the required phase advance between the two sextupoles. 



The favored solution to resolve the nonlinearity is to use a high frequency passive corrugated structure \cite{bane2016}, to correct the second order curvature prior to compression\cite{penco2017passive, craievich2010passive}. When a high current electron beam goes through a corrugated structure, the resulting wakefield can be used to modify the longitudinal phase distribution of the beam \cite{fu2015demonstration}. We considered the use of rectangular corrugated structures instead of round pipes, as this would allow one to vary the gap using a mechanical jaw adjustment. In addition, the effects on the emittance due to the transverse wakefields can be canceled using two flat corrugated structures rotated by 90 degrees with respect to each other \cite{lu2016time}. 

Accurate expressions for the short range wakefield of a flat dechirper are available in the literature \cite{bane2012,bane2016}. As an initial design study we considered the fundamental mode of longitudinal wake field in a slab geometry corrugated structure \cite{zhang_slabcorrugated}:
\eq{W(z)=\left(\f{Z_0 c}{\pi a^2}\right)\f{\pi^2}{16}\cos kz}
where the wave vector $k$ depends on the corrugation period $p$, the height $h$ and spacing $g$ of the corrugation teeth, and the pipe radius $a$ as $k=\sqrt{\f{p}{a h g}}$. In order to keep compatibility with the linac operation for APS storage ring operation, the linearizer can be located between L1 and L2.
    
    
The linearizer parameters were optimized using \textsc{elegant} simulations in order to take into account higher order terms in the longitudinal beam dynamics such as space charge forces, CSR and the small contributions from the vertical ramp and the injection chicane before the TESSA-266 setup. A global optimization selects the optimum aperture and wavevector as well as L2 phase to linearize the compression. The L2 voltage is then adjusted to keep constant energy in the chicane. Figure \ref{fig::lpswake}b shows \textsc{elegant} output at the prebuncher entrance, for the case where a 0.2~m long linearizer of 2.6~mm gap, and wave vector 1700 m$^{-1}$ is placed after L1 structure. 
As a quality factor for the linearization process we consider the fraction of e-beam slices that fill an ideal rectangular current distribution of 1~kA and 90~$\mu$m which can be used as a proxy for the FEL efficiency in the system. Figure \ref{fig::lpswake} shows the output longitudinal phase spaces (LPS) from \textsc{elegant} at the prebuncher entrance for the case of linearizer off (left) and linearizer on (right). The first one is characterized by a long talk characteristic of non linear compression. The red dashed line shows the ideal 1~kA rectangular current distribution. In the linearizer on case, more than 80\% of the beam slices fit within desired 1~kA rectangular current distribution while only 60\% without a linearizer. The double-horn shape of the LPS distribution could be suppressed by applying further compensation to eliminate the third order terms \cite{NSudar2020}, but due to the limited space in the bunch compressor it would be challenging to do this while limiting the transverse emittance growth. 

The 6-dimensional distribution \textsc{elegant} output  (x,xp,y,yp,t,p) is then converted to 6-dimensional \textsc{genesis} input (X,PX,Y,PY,T,P) to complete the TESSA-266 start-to-end simulations. After adjusting the position of the beam centroid to maximize the overlap with the input seed laser pulse, optimized time-dependent simulations yield up to 3 \% efficiency, much higher than the 1 \% efficiency obtained from the non-linearized compression case which is characterized by a very long low-current beam tail. The main reason of still large reduction in output power with respect to the ideal case is the large energy spread of the final beam. If this could be reduced to the design levels, efficiency above 5\% could be obtained. Future studies could involve optimization of the undulator tapering profile considering the actual temporal profile. 

 \FloatBarrier
\fig{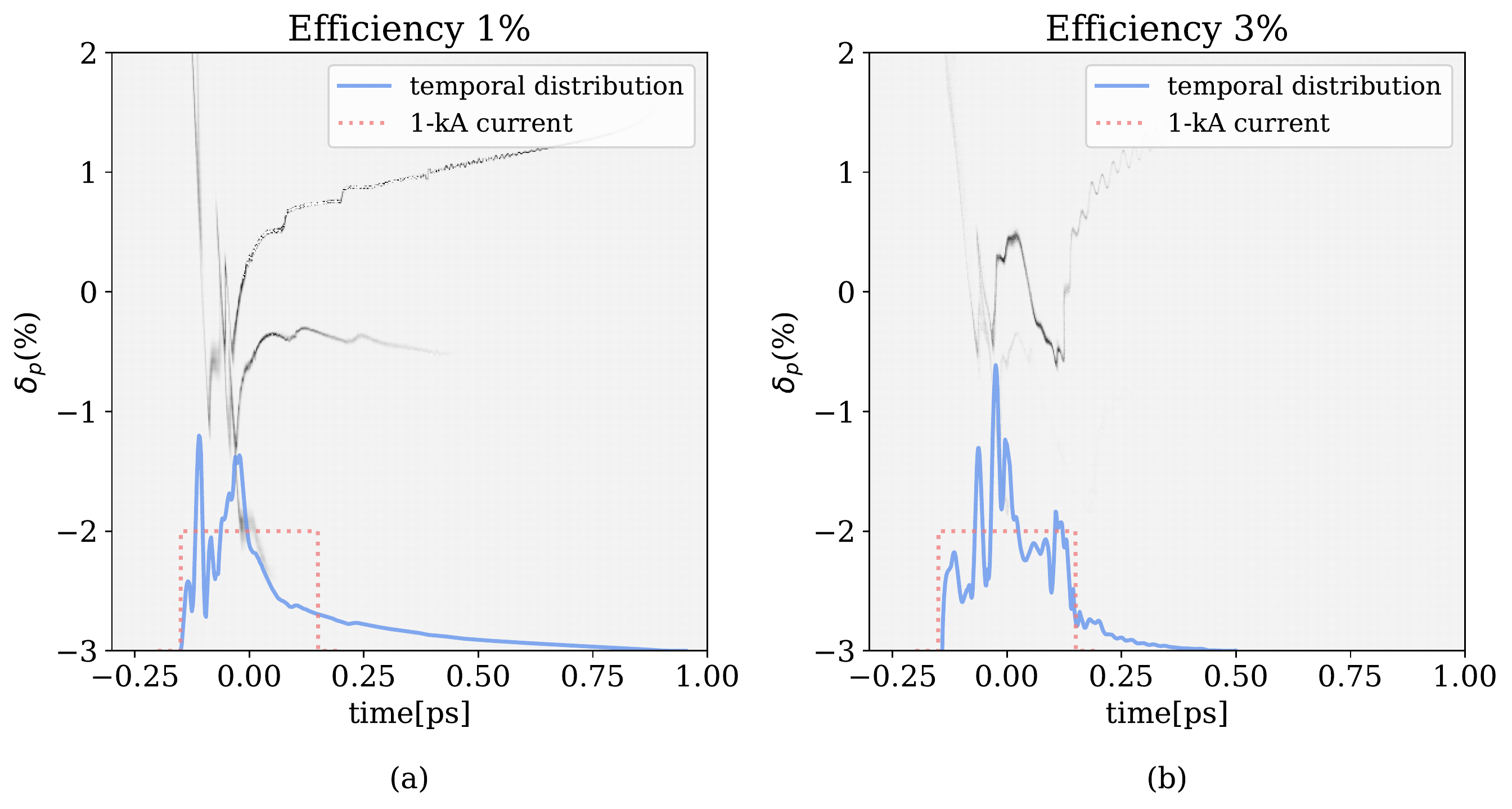}{Longitudinal phase space distribution (a) without (a) and with (b) passive corrugated waveguide linearizer. The projections on the time-axis are compared with the idealized 1~kA rectangular current distribution (red dashed line).\label{fig::lps} }{fig::lpswake}{1}

 \FloatBarrier
\pagebreak

\subsection{\label{Tolerances} Tolerance Study for the THESEUS system}

Once the initial design working point has been established and the tapering profile of the undulator defined, we carried out a study on the tolerances to key input parameters for the TESSA-266 system using  a series of \textsc{genesis} simulations. Acceptable parameters are defined as the ones that reduce the output power by less than 20 $\%$. It was found that when multiple parameters vary at the same time, at first order the reduction in output efficiency can be obtained simply as the quadrature sum of the contributions from each independent parameter variation off the ideal conditions.  In Table \ref{tab::jitter} we summarize the tolerance intervals obtained for all of the parameters including laser waist position, Rayleigh range, energy spread and transverse emittance.

These results are important not only to give an idea of the beam stability required for the experiment, but also as they allow to specify the level of accuracy of the various diagnostics used to characterize the TESSA-266 input beams.

\mt{c | c  c}{
\multicolumn{3}{c}{\textbf{Tolerance / Jitter Studies} }\\
\toprule
Parameter & Unit & Value/tolerance \\
\midrule
Relative magnetic errors  & \% &  $\pm$ 0.1    \\
Beam energy jitter & \% & $\pm$0.2 \\
Undulator section alignment & $\mu$m  & $\pm$60  \\
Time of arrival &  ps & $\pm$0.35 \\
\hline
Minimum input seed power & GW & 0.5 \\
Energy spread & \%  & $<$0.15\\
Transv. emittance & mm-mrad  & $<$2.5 \\
\bottomrule
}{tab::jitter}{Tolerance and jitter studies of key parameters in TESSA-266. The values represent the deviation from the nominal parameters which would be acceptable for the experiment.}

Fig. \ref{fig::tolerance}a shows the relative output power variation as a function of the relative seed laser time of arrival as estimated by performing multiple time-dependent simulations varying the seed pulse injection time compared to the peak of the e-beam temporal distribution. The result is $\pm$0.35 ps interval is attributable to the pulse length of the seed laser. The tolerance interval on the injection energy and the magnetic errors in the undulator (also shown in Fig. \ref{fig::tolerance}b,c) is set by the amplitude of the ponderomotive resonant bucket in the longitudinal phase space which is 0.2 $\%$. The energy stability of the APS linac (0.1 \%) should comfortably be within this tolerances. More attention will be needed to control the linac phase fluctuations to keep the time-of-arrival within the specified tolerances. Finally, in order to quantify the alignment error tolerances we used the simulation to predict the effects due to changes in electron beam position at each undulator section entrance. The results are shown in Fig. \ref{fig::tolerance}d.

\fig{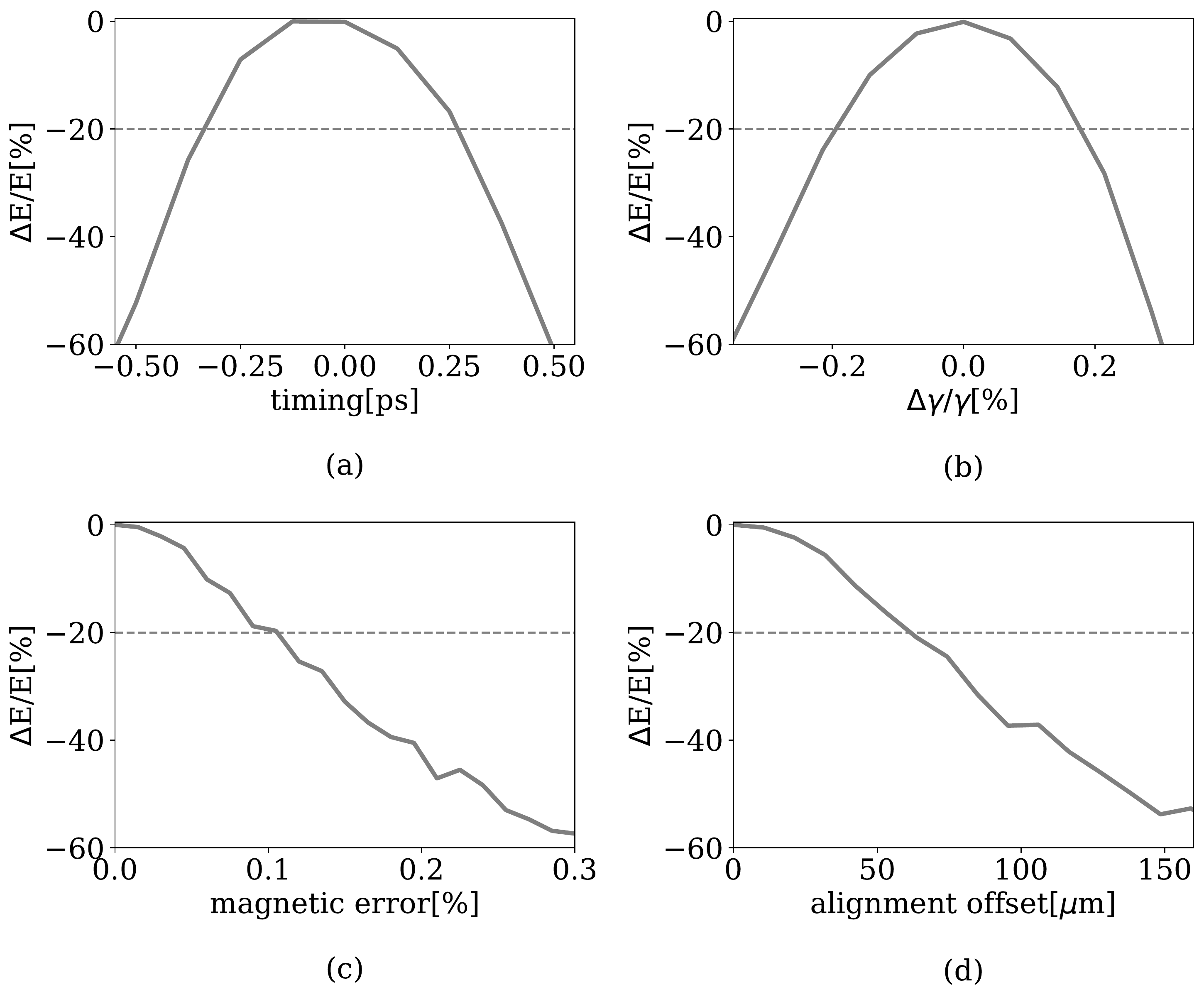}{Tolerance studies of TESSA-266 simulation. Output power variation as a function of (a) laser timing (b) energy jitter (c) magnetic errors in the undulator and (d) alignment tolerance as computed by \textsc{genesis} simulations with optimized tapering.}{fig::tolerance}{0.8}

\pagebreak

\section{\label{Undulator} Undulator construction, tuning and measurements}

\begin{figure}[ht!]
\centering
\includegraphics[clip,width=0.8\textwidth,page=1,trim=1.5in 0.3in 1.5in 0.3in]{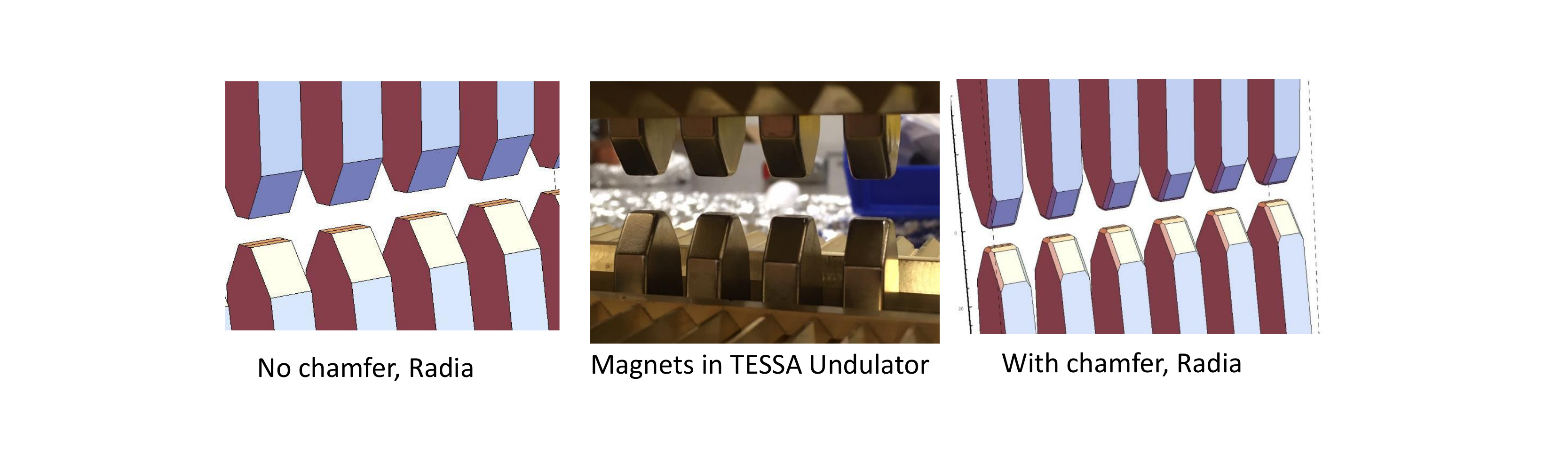}
\caption{left) \textsc{radia} simulation model using straight edge magnets. center) A picture of the magnets assembled in one of the THESEUS undulator sections. right) \textsc{radia} simulation model with chamfered magnets \label{fig::chamfer}}
\end{figure}

This section contains a more technical discussion of the undulator system design and construction. In order to determine the optimal undulator period length, we started by 3D magnetostatic simulations with \textsc{radia} \cite{radia} to calculate the magnetic field of two Halbach arrays \cite{halbach} of NdFeB permanent magnets (residual magnetization 1.18~T) rotated and shifted with respect to each other by 90 degrees. The undulator gap is determined as a result of a compromise between different demands. A smaller gap would allow to use a shorter undulator period which in principle increases the final output power, but the tighter aperture has the drawback of a smaller transverse acceptance and larger wakefield effects for the experiment.
It was found that small features on the magnets such as the chamfering of the edges would make an appreciable impact on the on-axis magnetic field.  In Fig. \ref{fig::kvslamb}, we plot the undulator factor $K$ as a function of the undulator period from magnetostatic \textsc{radia} simulations \cite{radia} for chamfered and unchamfered magnets for various gaps. The TESSA-266 FEL resonant condition $K^2 = \frac{2 \lambda_r}{\lambda_u} \gamma_r^2 -1$ is also shown. The design working point uses an undulator $K$ = 2.5 at a nominal gap of 5.8 mm.  Fig. \ref{fig::chamfer} better illustrates the mechanical differences between the original design using magnets with sharp edges and the more accurate simulation with chamfered magnets. 

\begin{table}[!hbt]
	\centering
\begin{threeparttable}
	\begin{tabular}{l|c|c}
	undulator parameters & unit & value \\
	\hline
	Type & & helical, permanent magnets\\
	Period length & mm & 32  \\
	Undulator length & $\lambda_u $ & 28 \\
	Magnetic field & T & 0.67-0.83 T \\
	Normalized vector potential, K & & 2-2.5 \\
    Magnetic gap \tnote{a} & mm &  5.58 - 7.54\\
    Magnet type &&  NdFeb\\
    Remnant field & T & 1.18\\
    \midrule
    \end{tabular}
    \caption{THESEUS undulator parameters}
	\label{tab:und}
	   \begin{tablenotes}
      \item[a]{the gap can be further increased by adding shims underneath the tuning plates}
    \end{tablenotes}
        \end{threeparttable}
\end{table}

\fig{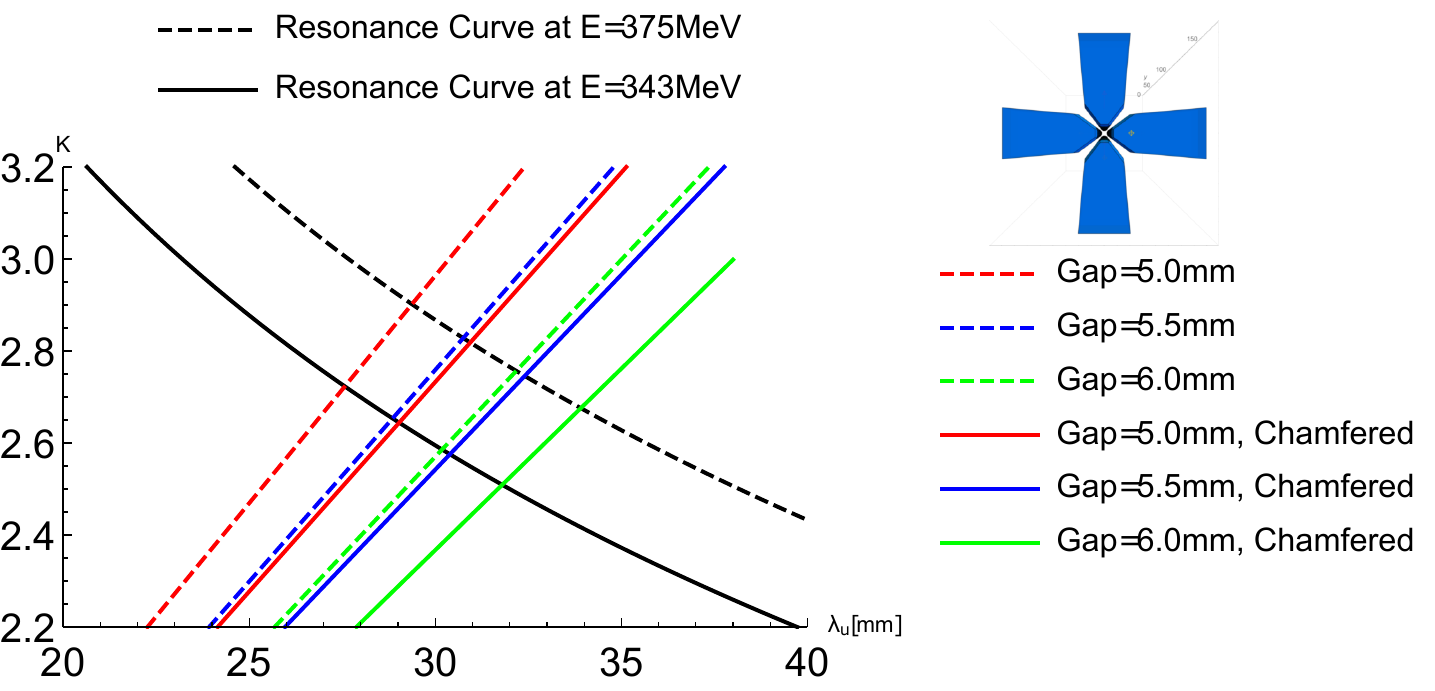}{Resonance Curve (black) and $K$ from \textsc{radia} simulation of the helical undulator plotted at different gaps. While original design had higher $K$ and beam energy (dashed), magnet chamfering resulted in lower $K$ (solid). In the inset a view from the beam axis of the permanent magnets used in the undulator construction is shown. }{fig::kvslamb}{0.8}

The undulator entrance and exit are designed to keep the center of the helical trajectory close to the beam axis thus ensuring strong overlap with the laser pulse. By comparing the outputs of period averaged codes and particle tracking simulations \cite{fisher:GPTFEL}, the lengths of the entrance and exit section add a 0.5 period of effective interaction, so that the effective length of the undulator in period-averaged \textsc{genesis} \cite{genesis} simulations discussed below is increased to 29 periods. 

The undulator section length is determined as a compromise between system performances and mechanical considerations. Shorter undulator sections allow less distance between the focusing elements and tighter beams while using longer sections maximizes the filling fraction for a given total system length and enables higher conversion efficiencies. At the same time, each undulator section length is limited by the size of the machining plane and practical size and weight considerations to $<$ 1 m. The final number of undulator sections was determined using the results of numerical simulations which indicated that 4 m of interaction length would enable approaching 10 $\%$ power conversion efficiency. It is useful to note here that a longer undulator system with additional sections would allow to reach even higher conversion efficiencies. 

The undulator vacuum chamber is made of non magnetic stainless steel and has wall thickness of 0.5 mm and average rms surface roughness of 1.6 $\mu$m. It is also divided in 4 different parts, each one of them 989.33 mm long and to be connected by CF133 flanges to the undulator break section chambers. A summary of the THESEUS undulator design parameters is presented in Table \ref{tab:und}.

To complete the discussion of the undulator system, we note how the length of the inter-undulator break section is a critical parameter as it will be discussed more in detail in Section \ref{Sect:Break-section}. For the present discussion, it is sufficient to say that diffraction dominates the evolution of the radiation in these sections where no energy exchange with the electron beam occurs, so it is important to minimize their length. At the same time, mechanical considerations (to allow enough space for all the different components to be housed in the break sections) put a limit on how short these can be.  Sophisticated engineering of the break section (permanent magnet quadrupoles, integrated phase shifter and both beam and radiation diagnostics) allowed to limit the length to 225~mm ensuring minimal diffraction/dephasing and strong continued interaction.

\subsection{Mechanical design and construction}

The undulator design is an evolution of the first tapered helical undulator built by UCLA for IFEL experiments \cite{duris2014high}. The undulator permanent magnets are designed with narrow, trapezoidal shape in order to increase field strength in the center of the gap, and the sides are slanted in order to prevent the magnets from slipping out of the holder. Each magnet is glued by Loctite M-31CL to an aluminum holder. Four brass strongbacks are fixed to the undulator end plates and are machined with v-shaped slots where the magnet holders can be inserted and slide. Each four sets of magnets are attached from the bottom of the holders to a tuning plate with \#6-32 screws which can be used to adjust the undulator tapering profile. This operation does require a re-measurement of the undulator trajectory so will not be possible once the undulator are installed on the beamline. Figure \ref{fig::assembly} shows a cross section view of the undulator technical drawing \cite{tara:napac}.

\fig{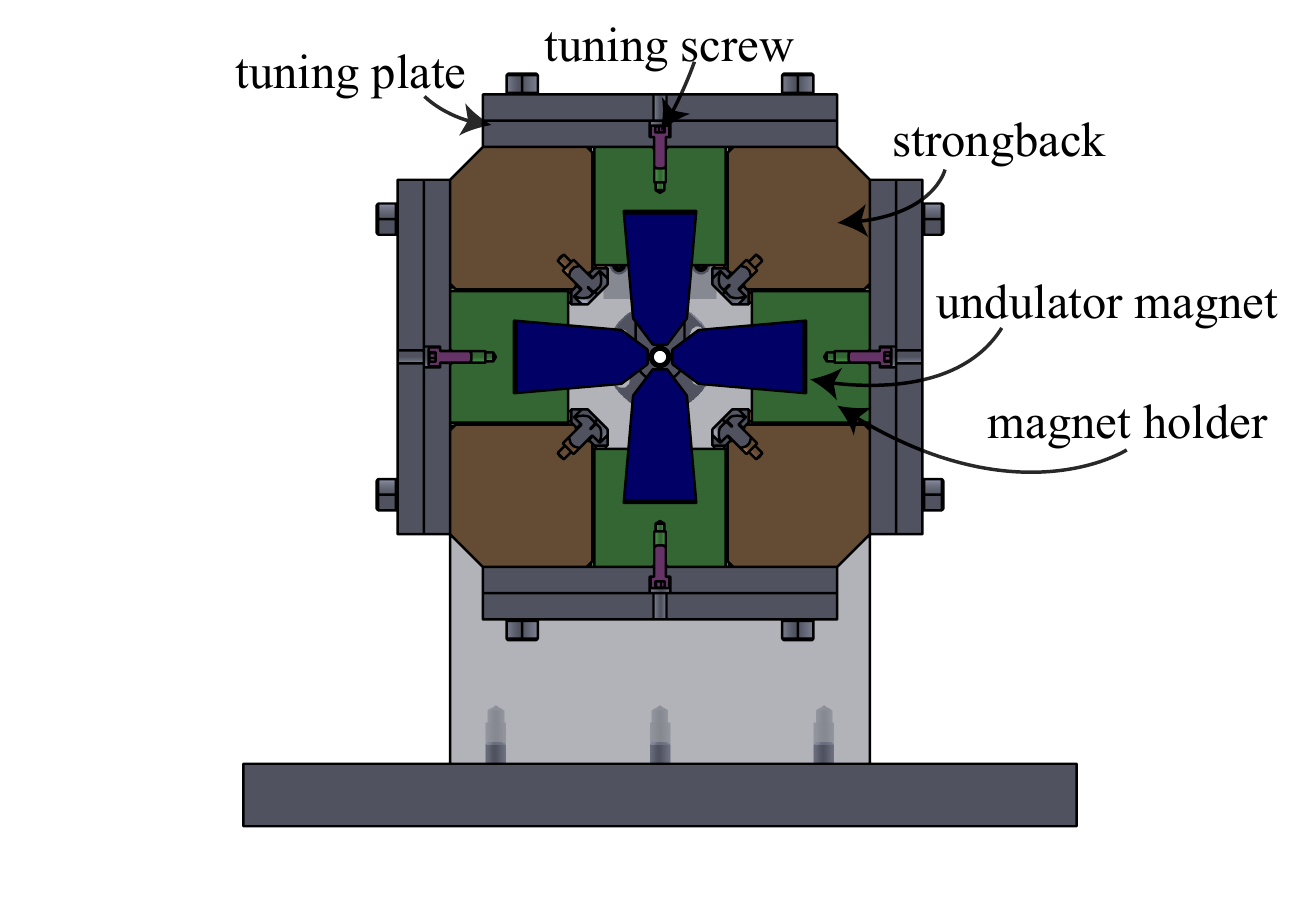}{A sectional view of undulator design showing 4 permanent magnets, their holders, tuning plates and and the 4 strong backs. The undulator support is also shown.}{fig::assembly}{0.8}

The undulator magnets are responsible for a non negligible magnetic force on the undulator structure. Based on \textsc{radia} simulations, there is inward force of up to 100N inward on the holders of the magnets with in-out magnetization and a net 46N after averaging on each undulator tuning plate. The total net inward force imposed on each strongback fixed on both ends is estimated to cause bending by 0.06mm based on Solidworks analysis.

\begin{figure}
  \begin{subfigure}[b]{0.33\columnwidth}
    \includegraphics[width=\linewidth]{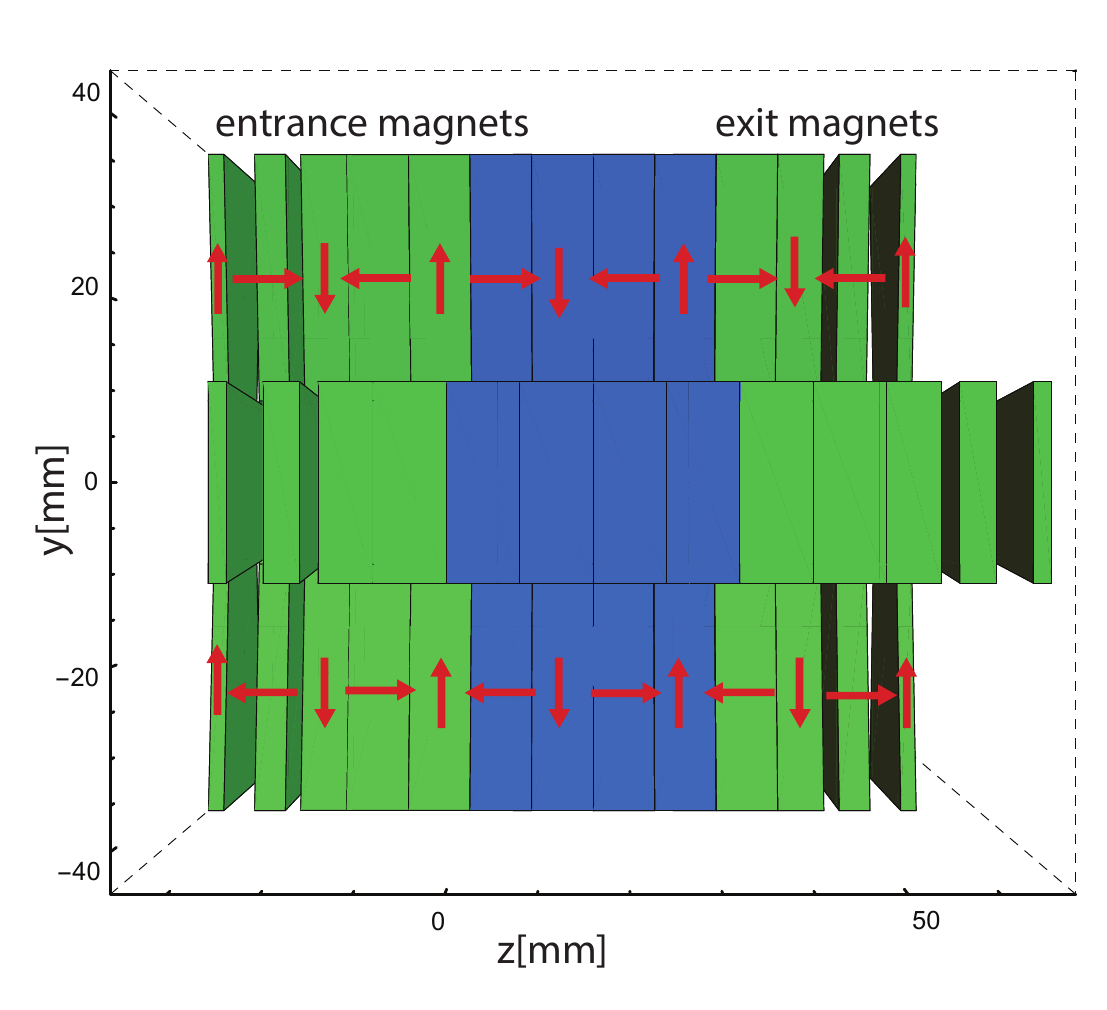}
    \caption{}
    \label{fig::ent}
  \end{subfigure}
  \hfill 
  \begin{subfigure}[b]{0.55\columnwidth}
    \includegraphics[width=\linewidth]{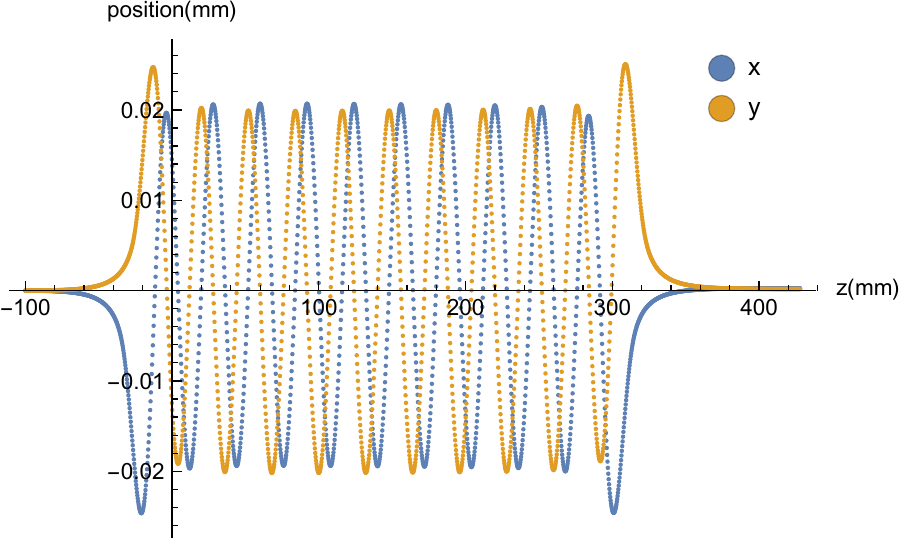}
    \caption{}
    \label{fig::traj}
  \end{subfigure}
  \caption{(a) Illustration of magnetization in entrance and exit sections (green) showing 2~mm, 4~mm, 8~mm, 8~mm magnets from left to the right on the entrance side. (b) e-beam trajectory in \textsc{radia} simulation with entrance and exit magnets properly tuned.}
  \label{fig::ent_tra}
\end{figure}

The entrance and exit sections are designed to minimize trajectory offset from the geometric center of the undulator. They use 2~mm, 4~mm, and two 8~mm magnets at each end of the array (see Fig. \ref{fig::ent}) \cite{Clarke} adding a physical length of 34mm at each end of the undulator. Special holders were used to hold the thinner magnets in the strongback. 

\begin{figure}[h!]
\centering
\includegraphics[width=0.9\linewidth]{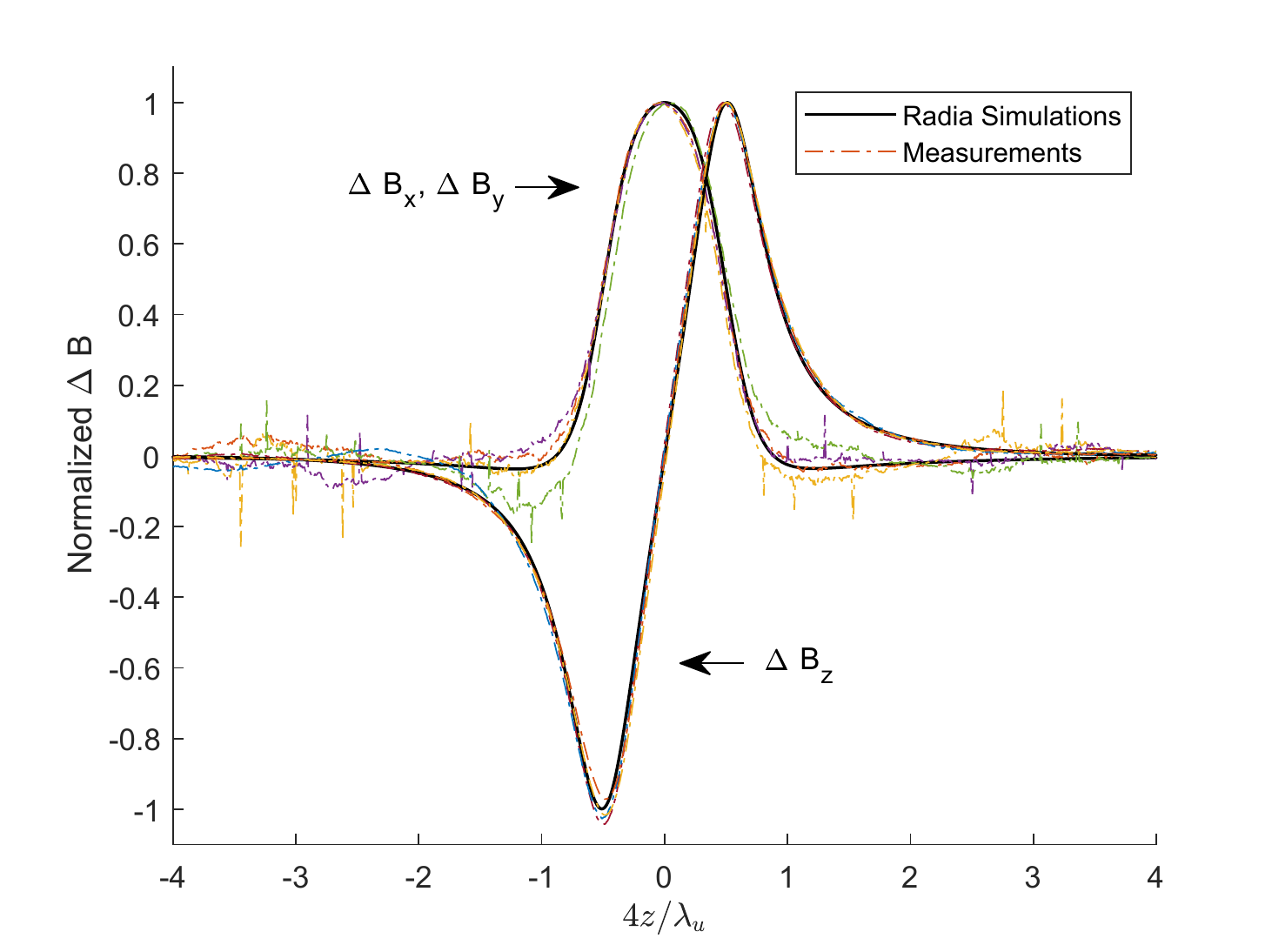}
  \caption{Comparison between \textsc{radia} simulation (solid line) and measurements (color lines) for the transverse and longitudinal eigenfunctions. The data is obtained simply measuring (or simulating) the effects of a shift of a single permanent magnet on the three magnetic field components.}  
  \label{fig:eigenfunctions}
\end{figure}

\subsection{Magnetic measurements and tuning Procedure}

The magnetic measurements are performed using a setup designed to precisely hold a narrow three axis Hall probe and guide it along the geometrical center of the assembly. In order to satisfy the tight requirements on the magnetic field errors along the undulator, positioning of the Hall probe should be within 200 $\mu$m of the beam axis with less than 2 degrees rotation error. The 2~mm wide Hall probe and the guiding mechanism are designed to fit inside the small diameter beam pipe. This is important as the relative magnetic permeability of the tube is specified to be less than 1.05, however maximum values of 1.08 were measured close at the weld joints. For this reason, final measurements and tuning of the magnetic field were performed with the pipe installed. 

The Hall probe scan yields the variation of the three field components $Bx, By, Bz$ as a function of $z$ and is the basic tool used to tune the magnetic field of the undulator. The lack of soft magnetic material in the THESEUS undulators allows the tuning procedure to rely on the principle of superposition. The change in field strength (transverse and longitudinal) as a function of $z$ when a single magnet is moved using a tuning screw is normalized and defined as the  magnet tuning eigenfunctions. The eigenfunctions are computed using \textsc{radia} and benchmarked against Hall probe measurements as shown in Fig. \ref{fig:eigenfunctions}. A \textsc{matlab} tuning code optimizes the linear combination of eigenfunctions that, when added to a first measurement scan, would minimize the error with respect to the target magnetic field profile. 



\begin{figure}[h!]
\centering
\begin{subfigure}[b]{.5\linewidth}
\includegraphics[scale=0.45]{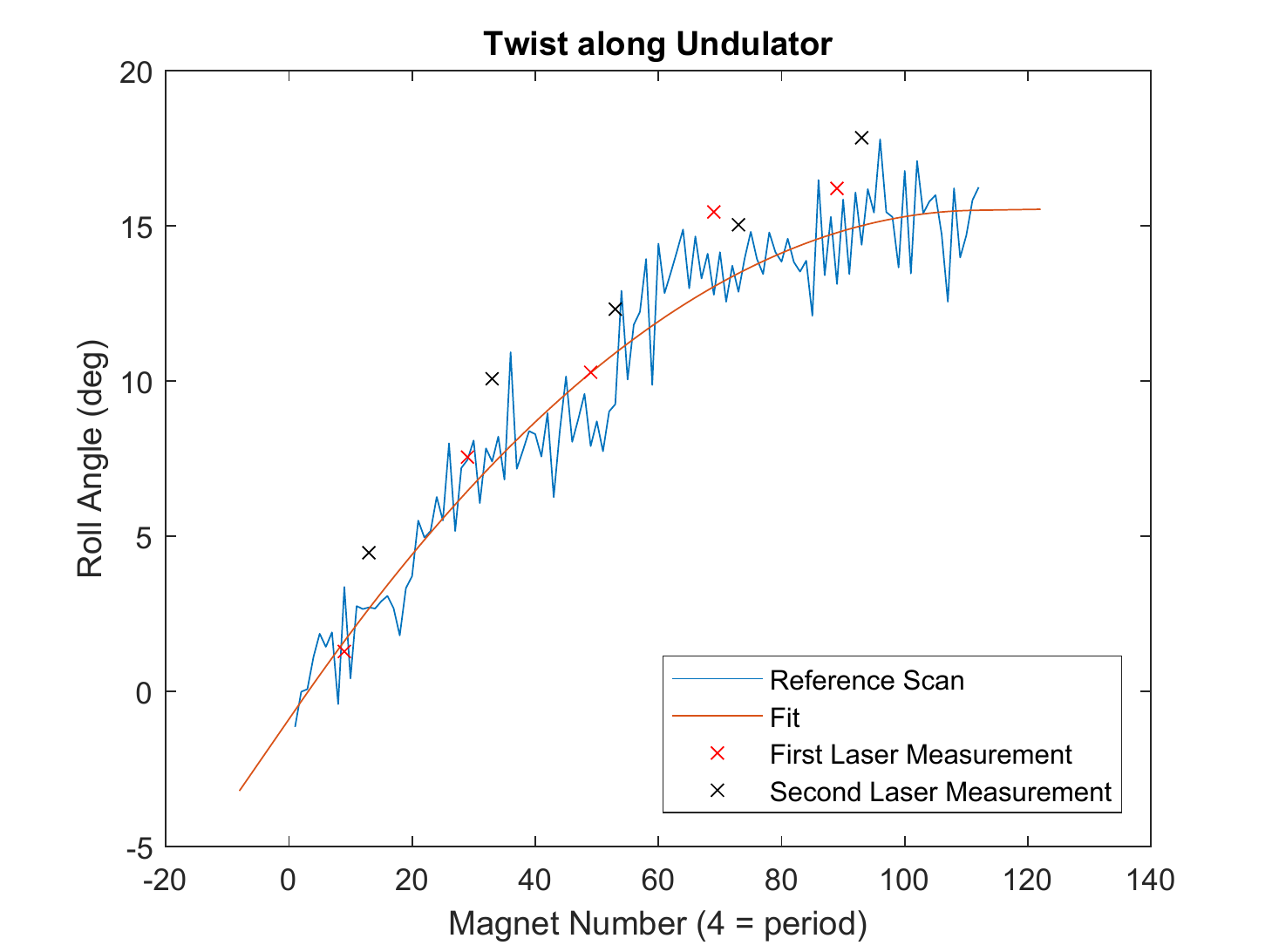}
\caption{}
\end{subfigure}%
\begin{subfigure}[b]{.5\linewidth}
\includegraphics[scale=0.45]{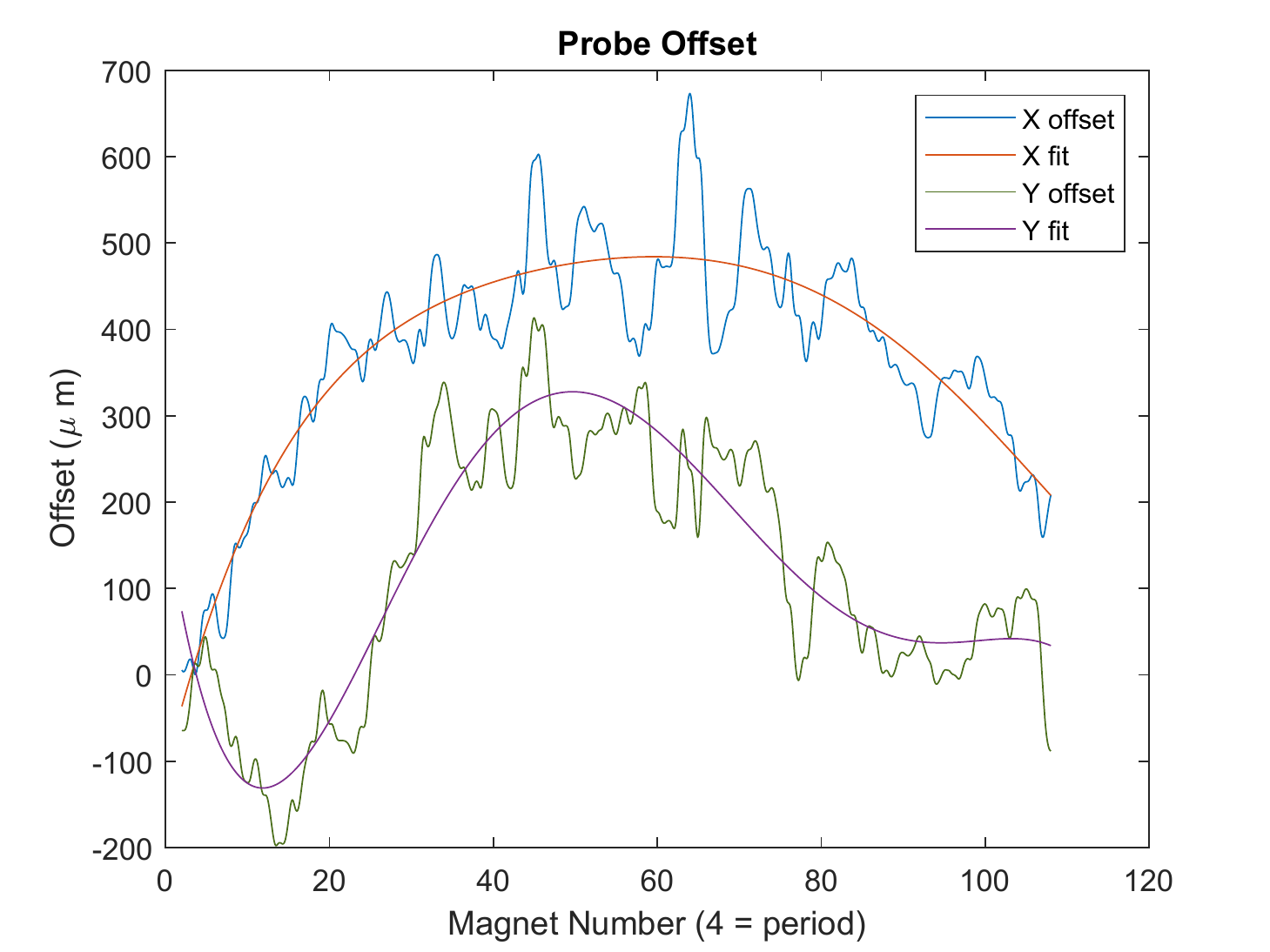}
\caption{}
\end{subfigure}
\caption{Measured Hall probe twist angle and transverse offset from ideal axis. This data is used to correct for the Hall probe position and angle and allow for magnetic field tuning within the required TESSA-266 tolerances.}
\label{fig:twistandoffset}
\end{figure}

The optimization fundamentally relies on the probe accurately measuring the fields along a fixed axis in the undulator. Systematic twisting or bending of the tuning carriage caused by machining errors or slight magnetic/gravitational forces must be understood and taken into account. Fortunately, these errors have predictable effects on measurements. Before tuning starts, a reference Hall probe scan is recorded with all magnets pulled out to their mechanical hardstops. An offset from the axis in the scan will result in a non zero measured $Bz$ field sinusoidal and in phase with the $Bx$ field ($y$ offset) or the $By$ field ($x$ offset). Additionally, a roll twist error will manifest in a shifting of the perceived peak field positions. The reference scan is used to generate a systematic transformation map which is applied to the data before optimization to account for such errors. The ability of the reference Hall probe scan to accurately compute probe position has been confirmed by directly observing the vacuum pipe bend (with one of the undulator arrays removed) under with effect of gravity with pulls in the $+\hat y$ direction (Figure \ref{fig:twistandoffset}b). The probe twist was consistent with independent measurements obtained reflecting a laser off a mirror glued above the probe carriage (Figure \ref{fig:twistandoffset}a).

\begin{figure}[h!]
\centering
\begin{subfigure}[b]{.5\linewidth}
\includegraphics[scale=0.45]{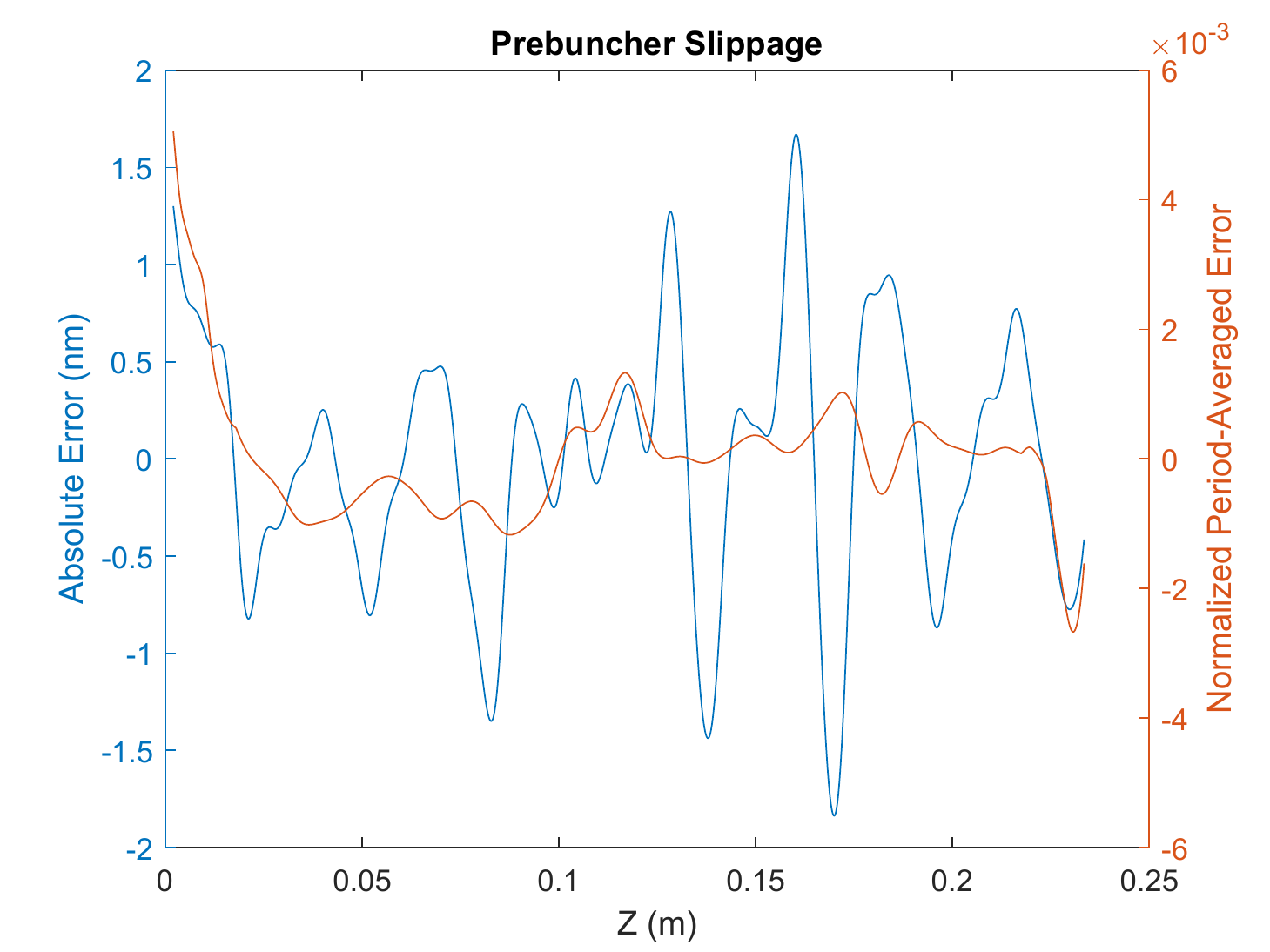}
\end{subfigure}%
\begin{subfigure}[b]{.5\linewidth}
\includegraphics[scale=0.45]{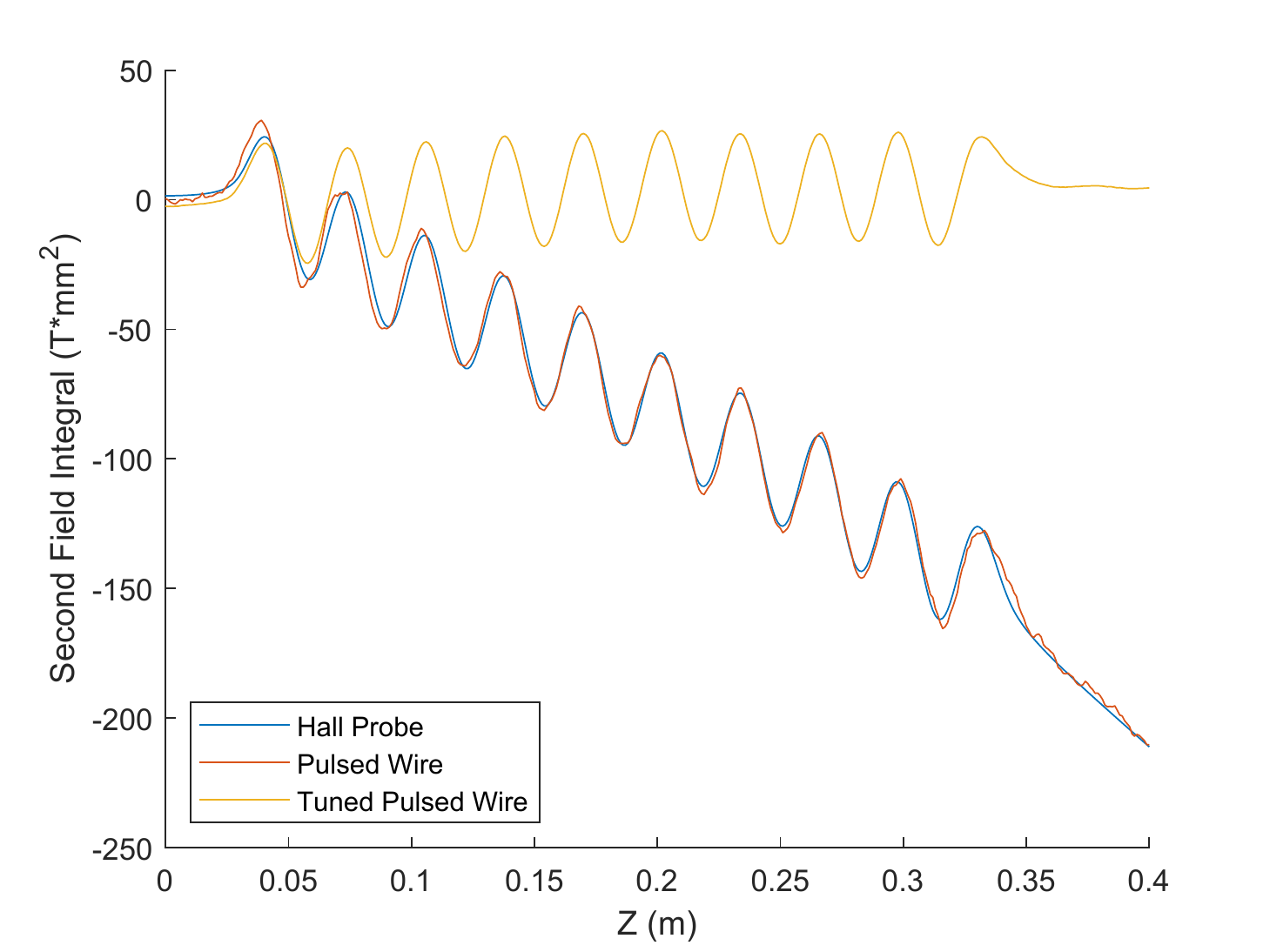}
\end{subfigure}
\caption{a) Tuned slippage error for the TESSA-266 prebuncher. b) Comparison of the By magnetic field component using hall probe and pulsed-wire measurements. The final pulse-wire measurement is done after tuning entrance and exit magnets.}
\label{fig:slippage}
\end{figure}

Finally, the fields are fine-tuned to minimize the beam slippage through the interaction as shown in Fig. \ref{fig:slippage}a to ensure that the particles see the same phase when they move through the different periods of the undulator. The magnetic error tolerance from the \textsc{genesis} simulations can be translated into a period-average slippage error of less than 0.5 \%. The picture show that the achieved tune satisfies the resonance requirement.



After the desired tapering profile is obtained with the Hall probe scans, entrance and exit magnets are tuned last using the pulsed-wire method to zero the first and second field integrals \cite{varfolomeev1994undulator, osmanov1998further, d2016ultra}. The method works by passing a current pulse through a wire tensioned through the undulator. The magnetic fields exert forces on the wire as the pulse passes through, creating wave vibrations that propagate in both directions. For a short pulse, the deflection of the wire is proportional to the first field integral and described the velocity of an electron passing through the undulator. For a long pulse, the deflection is proportional to the second field integral and describes an electrons trajectory. The main advantage of this method is in that it provides a direct measurement of the expected beam trajectory in the undulator and gives instantaneous feedback while tuning. Common limitations to be considered are the sag of the wire and dispersion in the vibration waves \cite{warren1988limitations, kumar2010analysis}. The beam trajectory before and after tuning the entrance and exit section for the prebuncher undulator are shown in Fig. \ref{fig:slippage}b.





\section{\label{Sect:Break-section} Break section design and doublet focusing}

\begin{figure}
    \includegraphics[width=\linewidth]{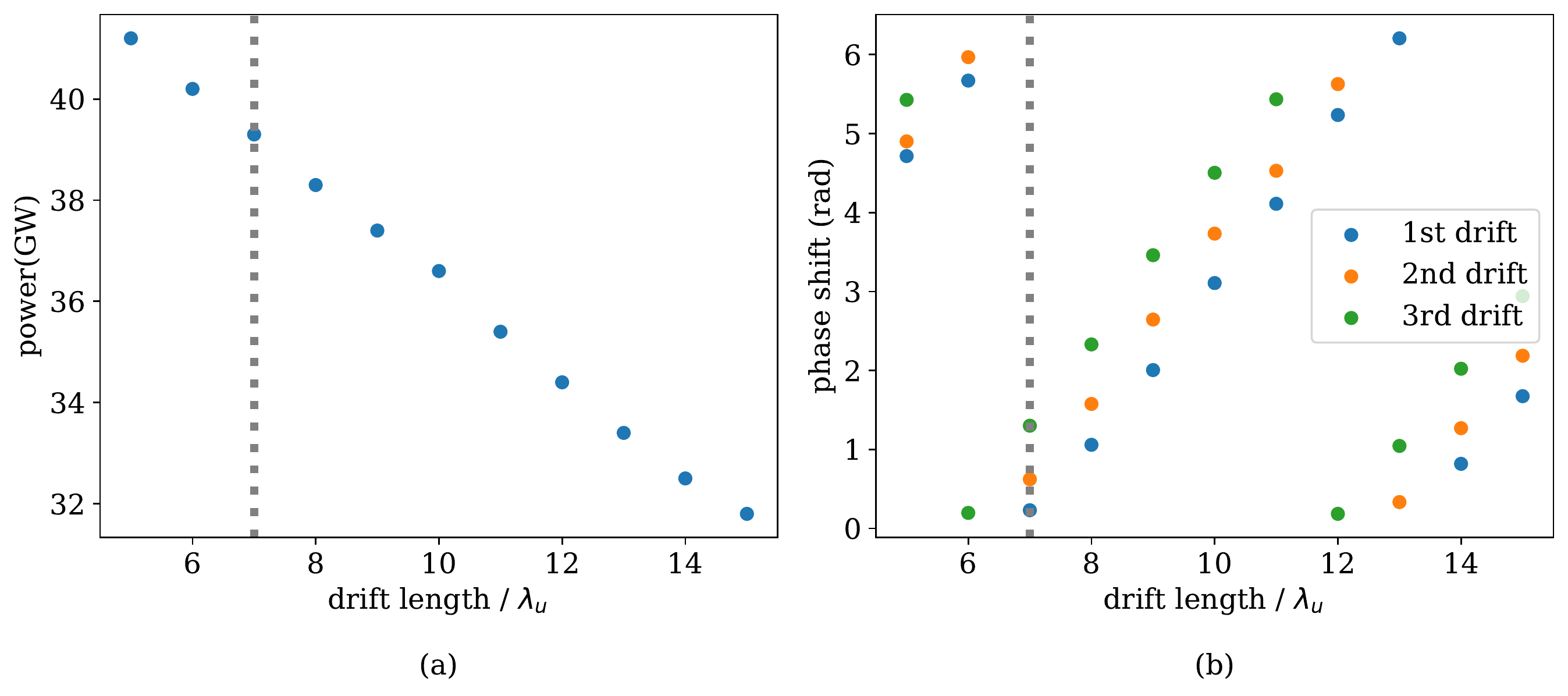}
  \caption{(a) Expected power output from the Genesis-Informed-Tapering optimization algorithm for different break section drift lengths in units of undulator period $\lambda_u$. (b) Optimal phase shift each undulator break section. The vertical dashed line indicates 7$\lambda_u$ drift length chosen for TESSA-266.\label{fig::powerdrift}}
\end{figure}

No description of the THESEUS undulator system would be complete without a detailed summary of the break section in between the undulator section. As mentioned in Section II, the length was determined as a compromise between engineering constraints and the observation that a shorter distance would lead to larger efficiency as shown in Figure \ref{fig::powerdrift}a. The output decrease is due to the fact that longer break sections allow the radiation to diffract quickly before the deceleration process is restarted in the next tapered undulator section, leading to particle detrapping and ultimately efficiency loss. 

In addition to the demand for the break section to be as short as possible, it is also imperative that the phase shift between the radiation and the electron is compensated to allow the particles to fall back in the ponderomotive bucket. In principle this happens automatically if the drift length is an integer multiple of $2\gamma^2\lambda_r$, where $\gamma$ is beam energy and $\lambda_r$ is laser wavelength. In practice, a phase shifter is required in a strongly tapered FEL system since the electron beam energy is changing each break section and one needs to compensate the phase shift due to the diffraction in the radiation field \cite{wu2018recent} to maximize the output power (see Fig. \ref{fig::powerdrift}a). In Fig. \ref{fig::powerdrift}b we show the optimal phase shifts for each undulator break section as a function of the drift length. 

\begin{figure}[h]
\centering
  \begin{subfigure}[b]{0.43\columnwidth}
    \includegraphics[width=\linewidth]{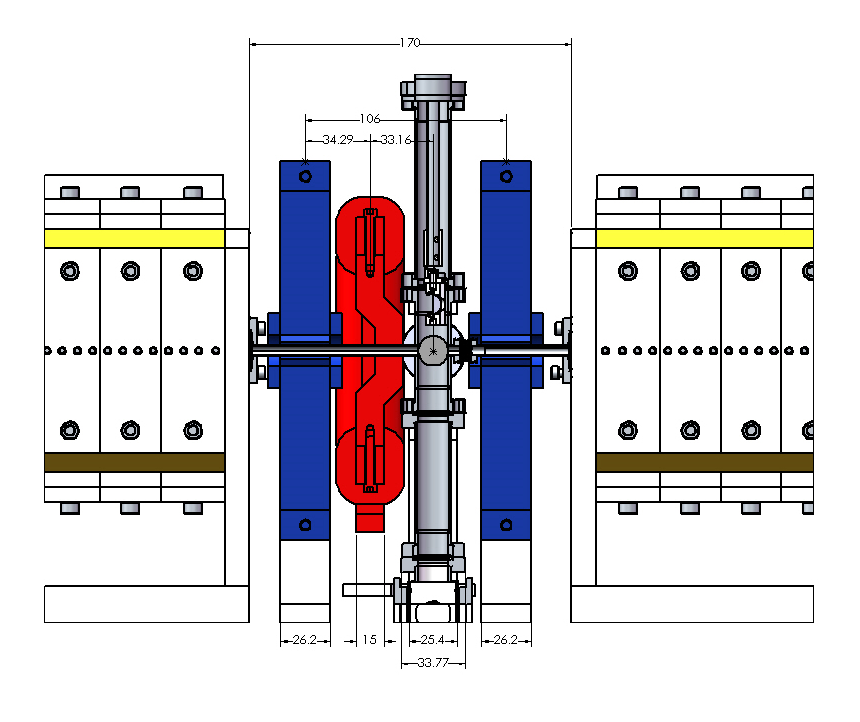}
    \caption{}
    \label{fig::breaksection}
  \end{subfigure}
  \hspace{1.2cm}
  \begin{subfigure}[b]{0.3\columnwidth}
    \includegraphics[width=\linewidth]{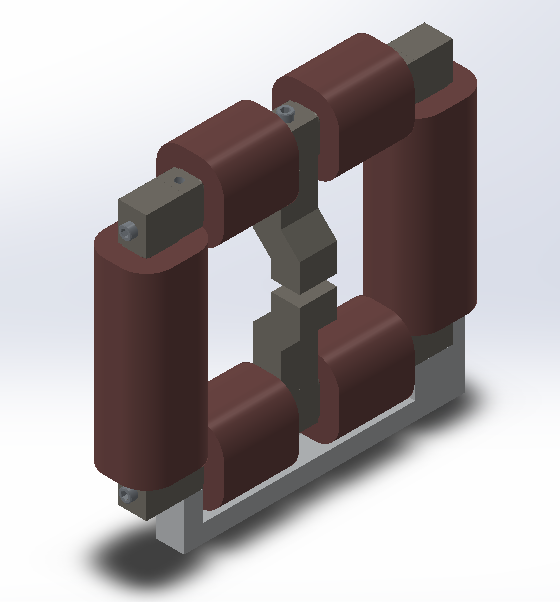}
    \caption{}
    \label{fig::emd}
  \end{subfigure}
  \caption{(a) Break section design of TESSA-266 with two hybrid permanent-magnet quadrupoles (red) and an electromagnetic dipole (blue). The dimensions shown in the figure are in mm. (b) H-shaped electromagnetic dipole with a protruded pole. }
  \label{fig::breakemd}
\end{figure}

In the final design, the break section of TESSA-266 contains a doublet of adjustable hybrid quadrupoles, an electromagnetic dipole, and diagnostics (Fig. \ref{fig::breaksection}) all compacted in a mechanical drift space of 170~mm. In \textsc{genesis} simulations (where due to the period average approximation entrance and exit sections can only be approximately included) we model the break sections with a 224 mm drift length (i.e. 7 $\lambda_u$). In order to control the phase shift of the electrons with the radiation, the two quadrupoles can be horizontally translated using remotely controlled translation stages to provide horizontal trajectory kicks and form together with the electromagnetic dipole in the center of the section a tunable phase shifter. In Fig. \ref{fig::emdplot}a,b we shows the required dipole magnetic field for each phase shift and the corresponding horizontal offsets for the quadrupoles. 


The electromagnetic dipole was designed with H-shaped coils and a protruded pole (towards the chicane center) as shown in Figure \ref{fig::emd} to provide maximum phase shift for equal dipole strength. The dipole magnetic length is 26~mm in the longitudinal dimension and has a tight transverse cross-section to avoid conflict with diagnostic ports. The expected saturation field of the magnet is 0.5T. The maximum phase shift that can be obtained is larger than  2$\pi$ at dipole strength of 0.25~T and quadrupole translation around 0.5~mm (Fig. \ref{fig::emd1}). 

The gradient of the quadrupole doublet is defined by the undulator parameter $K$ and the beam energy to satisfy the transverse matching of the beam in a periodic focusing channel. For the nominal tapering of $K$ from 2.5 to 2.1, corresponding integrated gradients would be around 5.62~T to 4.85~T. Since the length along the beamline is limited to $<$30 mm, these values corresponds to very high gradients in the range of 200 T/m. As a consequence we adopt the solution of permanent magnet quadrupoles, with the possibility of adjusting the gradient to optimize the output power. Note that the quadrupole gradients will not be remotely tunable, but having an adjustable gradient design simplifies tremendously fabrication tolerances. In addition, depending on when access to the experimental hall is allowed, it will be possible to hand-tune their focusing strength to optimize for power (for example squeezing the beam transversely in the last undulator section, or alternatively soften the focusing to better transport the decelerated electron bunches). We describe these novel hybrid quadrupoles in detail below.


\subfig{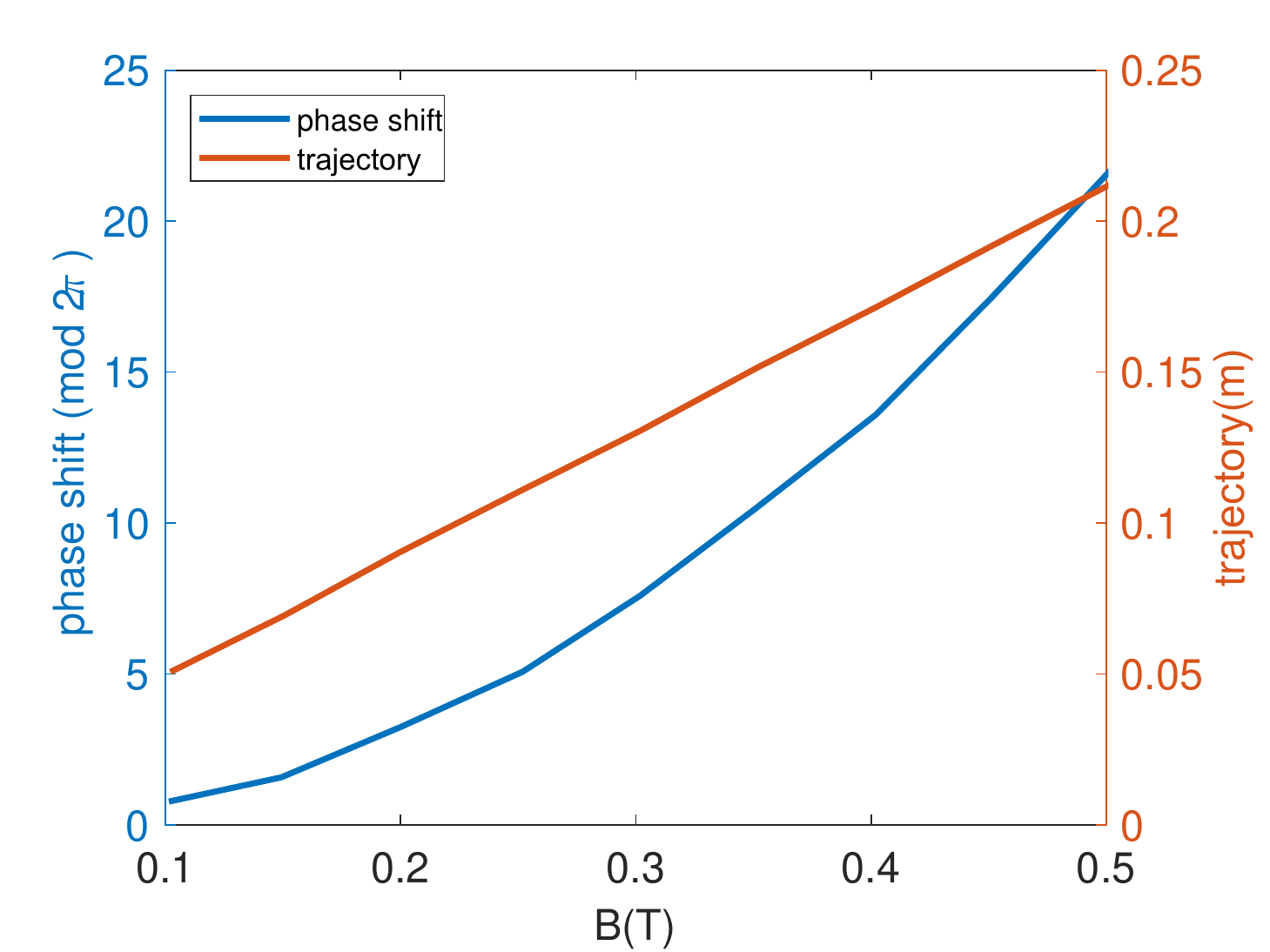}{fig::emd1}{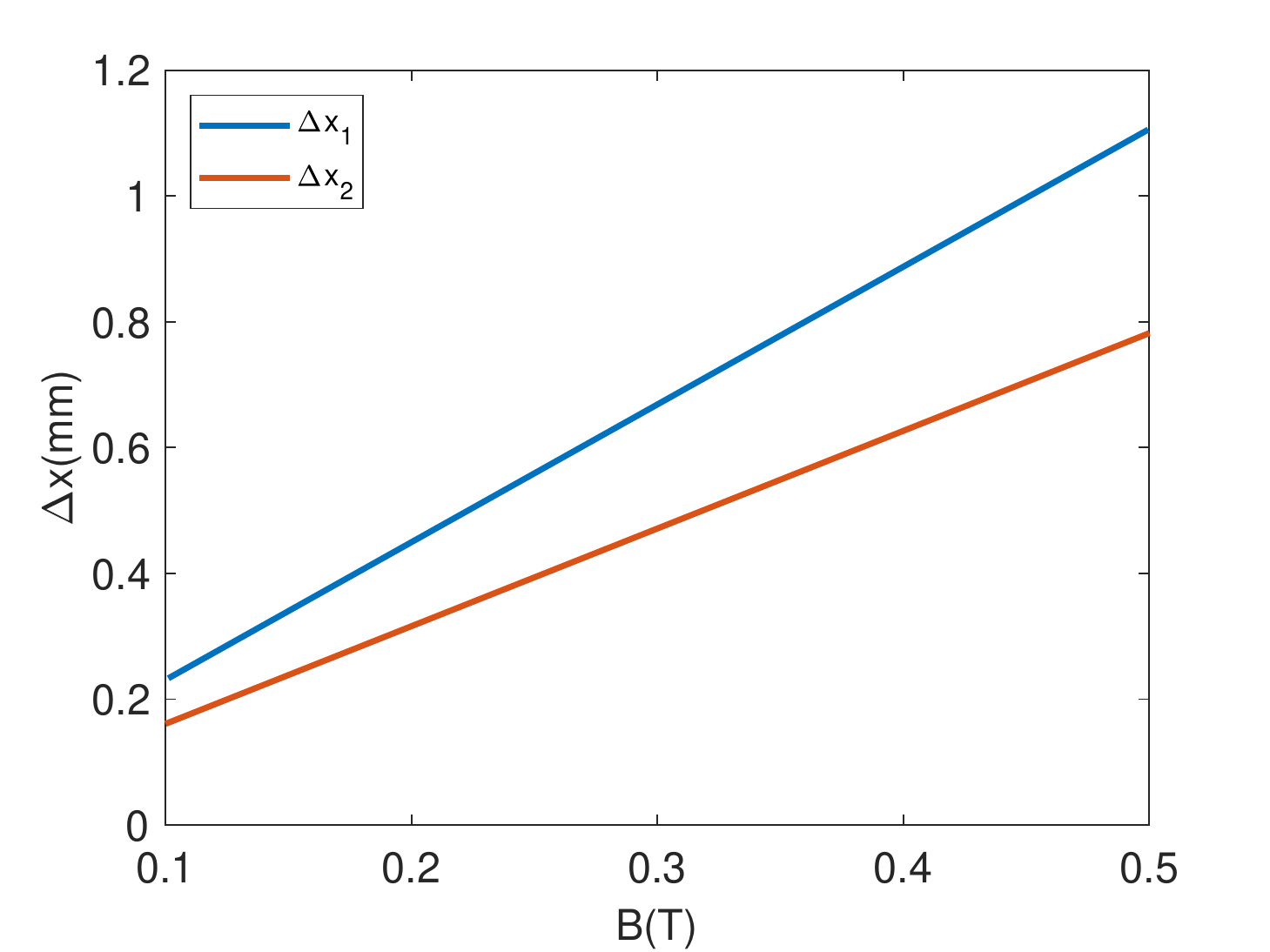}{fig::emd2}{(a) Phase shift corresponding to EMD field and (b) associated required horizontal shifts for the two permament magnet hybrid quadrupoles. In order to obtain a full 2$\pi$ phase shift the electromagnetic dipole field needs to provide 0.25T peak field and the quadrupoles are to be horizontally translated from their center by $\sim$ 0.5mm.}{fig::emdplot}

\subsection{\label{quadrupoles} Adjustable hybrid quadrupoles}
\begin{figure}[h!]
\centering
\includegraphics[scale=0.70]{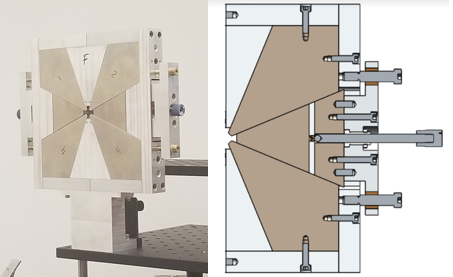}
  \caption{Completed Hybrid Quadrupole prototype and adjustable shim design.}
  \label{fig:hybridquad}
\end{figure}
The TESSA-266 Hybrid Quadrupole design consists of four steel poles with low carbon content ($<$0.06$\%$), and two NdFeB (1.45 T remanence) wedge sectors. Adjustable shims are used to create a magnetic circuit which can siphon magnetic flux density away from the aperture when the shims are brought in towards the magnet. The design has been chosen over other variable permanent quadrupole options \cite{halbach1983conceptual, marteau2017variable} to satisfy the constraints over the available beamline distance. When the shims are pulled away from the magnet, the magnetic circuit is severed, and the pole tips are maximally magnetized, still below the saturation limit. The shims are mounted to the steel poles and their positions are controlled with thumbscrews. A schematic is shown in Fig. \ref{fig:hybridquad}.

The pole tips are machined to match the shape of the ideal hyperbolic potential surfaces in the transverse plane having 4~mm pole tip radius relative to the symmetry axis. The magnetic design has been simulated in \textsc{radia} yielding a peak integrated gradient of 6.16~T, with an effective length of 30.9~mm with the shims all the way out and a range of tunability of up to 30$\%$ with a good field of region of at least 1.5 mm of radius. 
In that vicinity, the ratio of the quadrupole moment term to all other moments is small, i.e., $\frac{\int B_y(r=1.5mm,\theta)\cos(\theta)d\theta}{\int B_y(r=1.5mm,\theta)\cos(n\theta)d\theta}<0.001$. Simulation results also indicate magnetic center motion does remain within 50~$\mu$m of the geometric center when an error smaller than 1~mm is applied to the shim displacement position.

\begin{figure}[h!]
  \begin{subfigure}[b]{0.4\columnwidth}
    \includegraphics[width=\linewidth]{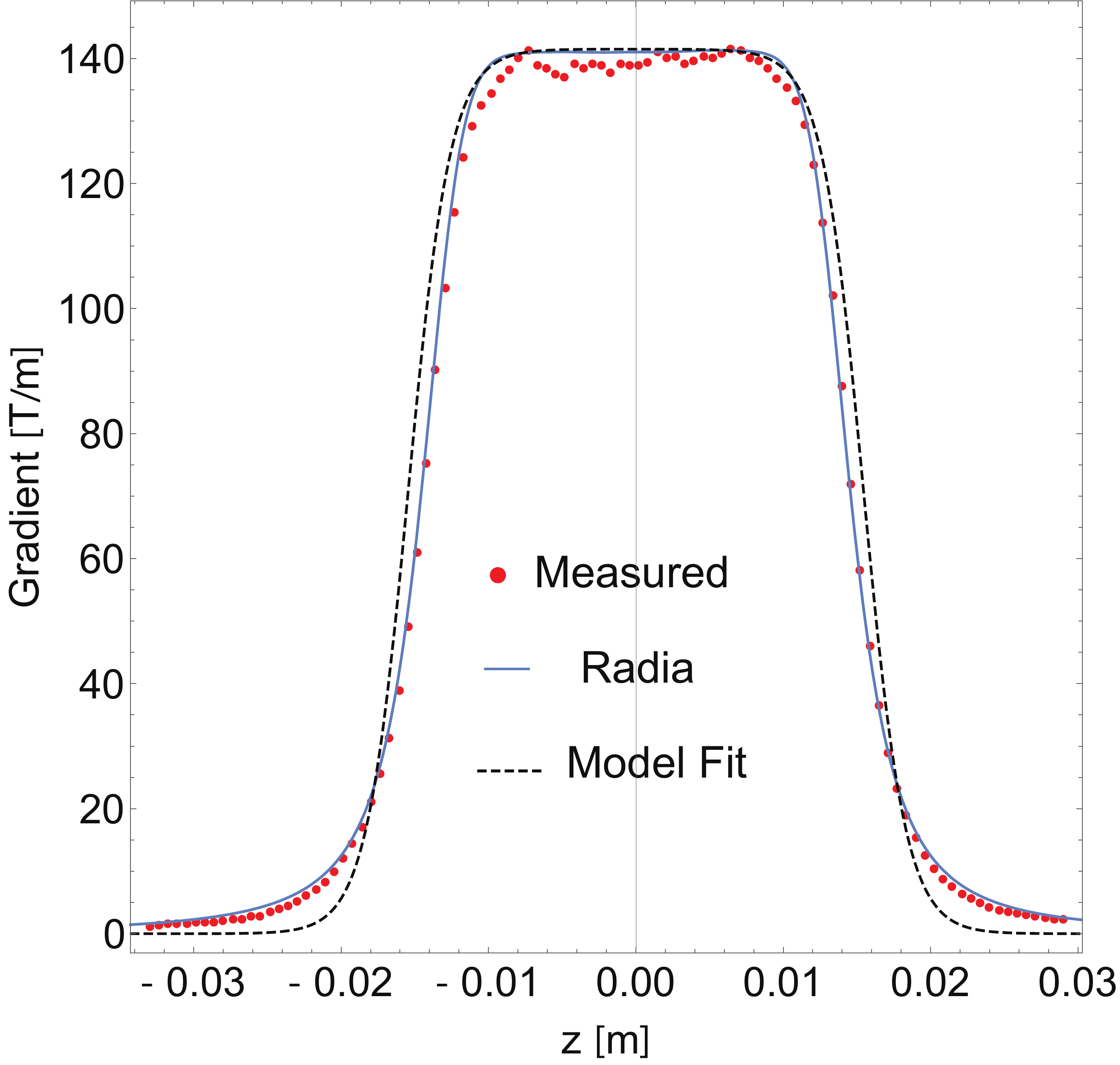}
    \caption{}
  \end{subfigure}
  \hfill 
  \begin{subfigure}[b]{0.45\columnwidth}
    \includegraphics[width=\linewidth]{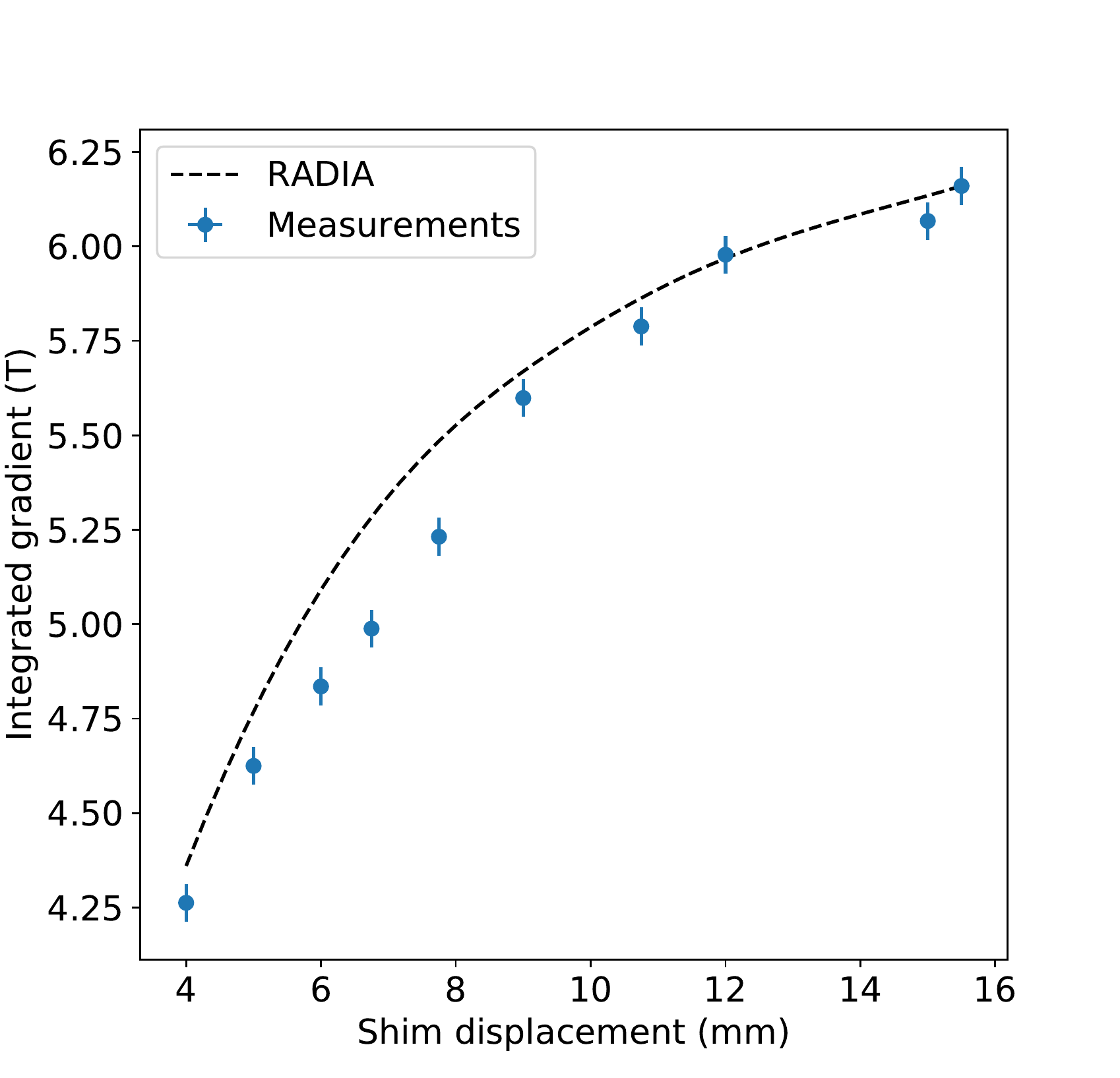}
    \caption{}
  \end{subfigure}
  \caption{(a) Measured longitudinal gradient profile compared with \textsc{radia} and model fit with shims all the way in. (b) Comparison of measured integrated gradient as a function of shim displacement and \textsc{radia} simulation.\label{hybridquadgrad}}
\end{figure}

A prototype hybrid quadrupole was constructed at RadiaBeam, and measured at UCLA, using an FW Bell STD18-0404 Hall probe. A Zaber translation stage was used to control the longitudinal motion of the probe. The motion is accurate down to 25~$\mu$m. Manual micrometers were used to control the transverse motion of the probe. We measured longitudinal gradient profiles for several shim displacements. In Fig. \ref{hybridquadgrad}(a), the measured gradient profile with shims all the way in is shown in comparison to the \textsc{radia} prediction assuming a magnetization of 1.39 T for the permanent magnets. The gradient longitudinal profile can be modeled reasonably well by:
\begin{equation}
G(z)=\frac{G_0(s)}{2}\left(\tanh\left(b z\right)-\tanh\left(b(z-L_\textrm{eff})\right)\right)
\end{equation}
where $G_0(s)$ is the peak gradient as a function of the shim displacement ($s$), $b$=350~m$^{-1}$ is the field stiffness parameter, and $L_\textrm{eff}$ as the effective length. The model fit to the measured data is also shown in Fig \ref{hybridquadgrad}(a). For this profile, the effective length of the quadrupole was found to be 30.7~mm $\pm$ 0.2~mm in agreement with the simulations. The integrated gradient as a function of shim displacement is shown in Fig. \ref{hybridquadgrad}(b). The integrated gradient was found to be adjustable from 4.2~T to 6.1~T fully satisfying the demands of the TESSA-266 design.  Finally, magnetic field was measured along the symmetry axis to estimate the magnetic center motion with a $<$ 50 $\mu$m resolution. The corresponding field variations were integrated and normalized by beam rigidity to give a 0.1~mrad bound on beam centroid kicks.

\section{\label{Conclusion} Conclusions}
In conclusion, in this paper we have presented the design of the tapered helical undulator system for the TESSA-266 experiment. The goals of the experiment are to obtain record high conversion efficiency in the UV region of the electromagnetic spectrum and provide experimental validation to the non linear TESSA regime of FELs where using an intense input seed and strongly tapered undulator, it is possible to overcome the instabilities that typically limit the energy extraction efficiency in single-pass FEL amplifiers. 

In order to minimize the electron beam size in the undulator as a means to achieve very high conversion efficiency, the undulator system uses a novel focusing scheme with a compact permanent magnet based quadrupole doublet in the undulator break sections. Start-to-end simulation studies highlight the importance of slippage effects in the interaction and the critical importance of controlling the longitudinal phase space evolution to maximize the conversion efficiency. 

While the UV spectral range is not particularly interesting as it is still within the reach of solid-state laser amplification and non linear optics, the improved understanding of the TESSA regime resulting from this experiment can be scaled to higher energy beams and shorter wavelength radiation sources. In fact, with higher beam brightness, shorter sections of tapered undulator including strong focusing schemes are already under study in major FEL facilities to improve the coupling of the relativistic beam with the radiation \cite{CEmma2015}. Another interesting direction to scale the results of the experiment will be to employ an electron bunch train (either in a warm linac, or even in an SRF accelerator) to increase the average power of the output radiation. In addition, if an electron bunch train were available, the possibility exist to re-utilize the TESSA-266 setup described here to study experimentally the start-up regime of a TESSA oscillator or TESSO \cite{TESSO} by embedding the undulator in an optical resonator. In this regime, a fraction of the amplifier output is redirected at the entrance of the system, offering a solution to the problem of intense seeding at wavelengths where commercial seed lasers are not available. 

\section{Acknowledgments}
We wish to acknowledge the support of DOE BES grant DE-SC0021190 and DOE SBIR/STTR grants DE-SC0017102, DE-SC0018559, and DE-SC0018571. This research used resources of the Advanced Photon Source, a U.S. Department of Energy (DOE) Office of Science User Facility, operated for the DOE Office of Science by Argonne National Laboratory under Contract No. DE-AC02-06CH11357. The work of WHT was supported by DOE award DE-SC0018656 to NIU. This research used resources of the National Energy Research Scientific Computing Center (NERSC), a U.S. Department of Energy Office of Science User Facility located at Lawrence Berkeley National Laboratory, operated under Contract No. DE-AC02-05CH11231 using NERSC award BES-ERCAP0019016.

\bibliographystyle{elsarticle-num} 

\bibliography{THESEUS_edit_v2}

\begin{thebibliography}{10}
\expandafter\ifx\csname url\endcsname\relax
  \def\url#1{\texttt{#1}}\fi
\expandafter\ifx\csname urlprefix\endcsname\relax\def\urlprefix{URL }\fi
\expandafter\ifx\csname href\endcsname\relax
  \def\href#1#2{#2} \def\path#1{#1}\fi

\bibitem{padamsee:SRF}
H.~Padamsee, 50 years of success for srf accelerators—a review,
  Superconductor science and technology 30~(5) (2017) 053003.

\bibitem{bane:NCRF}
K.~L. Bane, T.~L. Barklow, M.~Breidenbach, C.~P. Burkhart, E.~A. Fauve, A.~R.
  Gold, V.~Heloin, Z.~Li, E.~A. Nanni, M.~Nasr, et~al., An advanced ncrf linac
  concept for a high energy $e^+ e^-$ linear collider, SLAC-PUB-17301 (2018).

\bibitem{BPN}
R.~Bonifacio, C.~Pellegrini, L.~Narducci, Collective instabilities and
  high-gain regime in a free electron laser, Optics Communications 50~(6)
  (1984) 373--378.

\bibitem{KMR}
N.~Kroll, P.~Morton, M.~Rosenbluth, Free-electron lasers with variable
  parameter wigglers, IEEE Journal of Quantum Electronics 17~(8) (1981)
  1436--1468.

\bibitem{orzechowski}
T.~J. Orzechowski, B.~R. Anderson, J.~C. Clark, W.~M. Fawley, A.~C. Paul,
  D.~Prosnitz, E.~T. Scharlemann, S.~M. Yarema, D.~B. Hopkins, A.~M. Sessler,
  J.~S. Wurtele, High-efficiency extraction of microwave radiation from a
  tapered-wiggler free-electron laser, Physical Review Letters 57 (1986)
  2172--2175.

\bibitem{bonifacio1989tapering}
R.~Bonifacio, F.~Casagrande, M.~Ferrario, P.~Pierini, N.~Piovella, Tapering and
  self tapering in a free electron laser, in: High Gain, High Power Free
  Electron Laser: Physics and Application to Tev Particle Acceleration,
  Elsevier, 1989, pp. 227--242.

\bibitem{scharlemann1988selected}
E.~Scharlemann, Selected topics in fels, Tech. Rep. UCRL-99922, Lawrence
  Livermore National Lab., CA (USA) (1988).

\bibitem{feldman1989high}
D.~W. Feldman, H.~Takeda, R.~W. Warren, J.~E. Sollid, W.~E. Stein, W.~J.
  Johnson, A.~H. Lumpkin, R.~B. Feldman, High extraction efficiency experiments
  with the los alamos free electron laser, Nuclear Instruments and Methods in
  Physics Research Section A: Accelerators, Spectrometers, Detectors and
  Associated Equipment 285~(1-2) (1989) 11--16.

\bibitem{hafizi1990efficiency}
B.~Hafizi, A.~Ting, P.~Sprangle, C.~Tang, Efficiency enhancement and optical
  guiding in a tapered high-power finite-pulse free-electron laser, Physical
  Review Letters 64~(2) (1990) 180.

\bibitem{fawley:taperedsase}
W.~M. Fawley, Z.~Huang, K.-J. Kim, N.~A. Vinokurov, Tapered undulators for sase
  fels, Nuclear Instruments and Methods in Physics Research Section A:
  Accelerators, Spectrometers, Detectors and Associated Equipment 483~(1-2)
  (2002) 537--541.

\bibitem{jiao2012modeling}
Y.~Jiao, J.~Wu, Y.~Cai, A.~Chao, W.~Fawley, J.~Frisch, Z.~Huang, H.-D. Nuhn,
  C.~Pellegrini, S.~Reiche, Modeling and multidimensional optimization of a
  tapered free electron laser, Physical Review Special Topics-Accelerators and
  Beams 15~(5) (2012) 050704.

\bibitem{emma:tapering}
C.~Emma, K.~Fang, J.~Wu, C.~Pellegrini, High efficiency, multiterawatt x-ray
  free electron lasers, Physical Review Accelerators and Beams 19~(2) (2016)
  020705.

\bibitem{makcurbis}
A.~Mak, F.~Curbis, S.~Werin, Model-based optimization of tapered free-electron
  lasers, Physical Review Special Topics-Accelerators and Beams 18~(4) (2015)
  040702.

\bibitem{schneidmiller:tapering}
E.~A. Schneidmiller, M.~Yurkov, Optimization of a high efficiency free electron
  laser amplifier, Physical Review Special Topics-Accelerators and Beams 18~(3)
  (2015) 030705.

\bibitem{fratalocchi}
A.~Fratalocchi, G.~Ruocco, Single-molecule imaging with x-ray free-electron
  lasers: Dream or reality?, Physical Review Letters 106~(10) (2011) 105504.

\bibitem{schwinger:XFEL}
R.~Alkofer, M.~Hecht, C.~D. Roberts, S.~Schmidt, D.~Vinnik, Pair creation and
  an x-ray free electron laser, Physical Review Letters 87~(19) (2001) 193902.

\bibitem{pagani:EUVlitho}
C.~Pagani, E.~Saldin, E.~Schneidmiller, M.~Yurkov, Design considerations of
  10kw-scale extreme ultraviolet sase fel for lithography, Nuclear Instruments
  and Methods in Physics Research Section A: Accelerators, Spectrometers,
  Detectors and Associated Equipment 463~(1-2) (2001) 9--25.

\bibitem{hosler:EUVFEL}
E.~R. Hosler, O.~R. Wood~II, W.~A. Barletta, Free-electron laser emission
  architecture impact on extreme ultraviolet lithography, Journal of
  Micro/Nanolithography, MEMS, and MOEMS 16~(4) (2017) 041009.

\bibitem{murokh:EUVlitho}
A.~Murokh, P.~Musumeci, A.~Zholents, S.~Webb, Towards a compact high efficiency
  fel for industrial applications, in: Compact EUV \& X-ray Light Sources,
  Optical Society of America, 2020, pp. EW4A--3.

\bibitem{JDuris2015}
J.~Duris, A.~Murokh, P.~Musumeci, Tapering enhanced stimulated superradiant
  amplification, New Journal of Physics 17~(6) (2015) 063036.

\bibitem{gover:RMP}
A.~Gover, R.~Ianconescu, A.~Friedman, C.~Emma, N.~Sudar, P.~Musumeci,
  C.~Pellegrini, Superradiant and stimulated-superradiant emission of bunched
  electron beams, Reviews of Modern Physics 91~(3) (2019) 035003.

\bibitem{hajima:high_efficiency}
H.~Zen, H.~Ohgaki, R.~Hajima, High-extraction-efficiency operation of a
  midinfrared free electron laser enabled by dynamic cavity desynchronization,
  Physical Review Accelerators and Beams 23~(7) (2020) 070701.

\bibitem{Colsonsideband}
R.~A. Freedman, W.~Colson, The sideband instability in free electron laser
  oscillators: Effects of tapering and electron energy spread, Optics
  communications 52~(6) (1985) 409--414.

\bibitem{davidson1987single}
R.~C. Davidson, J.~S. Wurtele, Single-particle analysis of the free-electron
  laser sideband instability for primary electromagnetic wave with constant
  phase and slowly varying phase, The Physics of fluids 30~(2) (1987) 557--569.

\bibitem{tsai2017sideband}
C.-Y. Tsai, J.~Wu, C.~Yang, M.~Yoon, G.~Zhou, Sideband instability analysis
  based on a one-dimensional high-gain free electron laser model, Physical
  Review Accelerators and Beams 20~(12) (2017) 120702.

\bibitem{NSudar2016}
N.~Sudar, P.~Musumeci, J.~Duris, I.~Gadjev, M.~Polyanskiy, I.~Pogorelsky,
  M.~Fedurin, C.~Swinson, K.~Kusche, M.~Babzien, A.~Gover, High efficiency
  energy extraction from a relativistic electron beam in a strongly tapered
  undulator, Physical Review Letters 117 (2016) 174801.
\newblock \href {https://doi.org/10.1103/PhysRevLett.117.174801}
  {\path{doi:10.1103/PhysRevLett.117.174801}}.

\bibitem{musumeci2018advances}
P.~Musumeci, J.~G. Navarro, J.~Rosenzweig, L.~Cultrera, I.~Bazarov, J.~Maxson,
  S.~Karkare, H.~Padmore, Advances in bright electron sources, Nuclear
  Instruments and Methods in Physics Research Section A: Accelerators,
  Spectrometers, Detectors and Associated Equipment 907 (2018) 209--220.

\bibitem{emma2017high}
C.~Emma, N.~Sudar, P.~Musumeci, A.~Urbanowicz, C.~Pellegrini, High efficiency
  tapered free-electron lasers with a prebunched electron beam, Physical Review
  Accelerators and Beams 20~(11) (2017) 110701.

\bibitem{lewellen1999_apslinac}
J.~Lewellen, K.~Thompson, J.~Jagger, S.~Milton, A.~Nassiri, M.~Borland,
  D.~Mangra, A hot-spare injector for the aps linac, in: Proceedings of the
  1999 Particle Accelerator Conference (Cat. No. 99CH36366), Vol.~3, IEEE,
  1999, pp. 1979--1981.

\bibitem{Shin2018}
S.~Shin, Y.~Sun, J.~Dooling, M.~Borland, A.~Zholents,
  \href{https://link.aps.org/doi/10.1103/PhysRevAccelBeams.21.060101}{Interleaving
  lattice for the argonne advanced photon source linac}, Physical Review
  Accelerators and Beams 21 (2018) 060101.
\newblock \href {https://doi.org/10.1103/PhysRevAccelBeams.21.060101}
  {\path{doi:10.1103/PhysRevAccelBeams.21.060101}}.
\newline\urlprefix\url{https://link.aps.org/doi/10.1103/PhysRevAccelBeams.21.060101}

\bibitem{TESSO}
J.~Duris, P.~Musumeci, N.~Sudar, A.~Murokh, A.~Gover, Tapering enhanced
  stimulated superradiant oscillator, Physical Review Accelerators and Beams
  21~(8) (2018) 080705.

\bibitem{lumpkin2001first}
A.~Lumpkin, R.~Dejus, W.~Berg, M.~Borland, Y.~Chae, E.~Moog, N.~Sereno,
  B.~Yang, First observation of z-dependent electron-beam microbunching using
  coherent transition radiation, Physical Review Letters 86~(1) (2001) 79.

\bibitem{lumpkin2002evidence}
A.~Lumpkin, R.~Dejus, J.~Lewellen, W.~Berg, S.~Biedron, M.~Borland, Y.~Chae,
  M.~Erdmann, Z.~Huang, K.-J. Kim, et~al., Evidence for microbunching
  “sidebands” in a saturated free-electron laser using coherent optical
  transition radiation, Physical Review Letters 88~(23) (2002) 234801.

\bibitem{lumpkin2003transverse}
A.~Lumpkin, Y.~Chae, J.~Lewellen, W.~Berg, M.~Borland, S.~Biedron, R.~Dejus,
  M.~Erdmann, Z.~Huang, K.-J. Kim, et~al., Evidence for transverse dependencies
  in cotr and microbunching in a sase fel, Nuclear Instruments and Methods in
  Physics Research Section A: Accelerators, Spectrometers, Detectors and
  Associated Equipment 507~(1-2) (2003) 200--204.

\bibitem{lumpkin:FEL}
A.~Lumpkin, D.~Rule, Feasibility of single-shot microbunching diagnostics for a
  pre-bunched beam at 266 nm, in: Proceedings, 39th International Free Electron
  Laser Conference, FEL2019, 2019, p. WEP041.

\bibitem{lumpkin:IBIC}
A.~Lumpkin, et~al., Proposed research with microbunched beams at lea, in: Proc.
  of International Beam Instrumentaion Conference, 2021, p. TUPP19.

\bibitem{duris2012}
J.~Duris, P.~Musumeci, R.~Li, Inverse free electron laser accelerator for
  advanced light sources, Physical Review Special Topics - Accelerators and
  Beams 15~(6) (2012) 061301.
\newblock \href {https://doi.org/10.1103/PhysRevSTAB.15.061301}
  {\path{doi:10.1103/PhysRevSTAB.15.061301}}.

\bibitem{YPark2018}
Y.~Park, P.~Musumeci, N.~Sudar, A.~Zholents, A.~Murokh, Y.~Sun, S.~Webb,
  C.~Hall, D.~Bruhwiler, {Strongly tapered undulator design for high efficiency
  and high gain amplification at 266 nm}, in: {Proceedings, 38th International
  Free Electron Laser Conference, FEL2017}, 2018, p. MOP011.
\newblock \href {https://doi.org/10.18429/JACoW-FEL2017-MOP011}
  {\path{doi:10.18429/JACoW-FEL2017-MOP011}}.

\bibitem{milton2001exponential}
S.~Milton, E.~Gluskin, N.~Arnold, C.~Benson, W.~Berg, S.~Biedron, M.~Borland,
  Y.-C. Chae, R.~Dejus, P.~Den~Hartog, et~al., Exponential gain and saturation
  of a self-amplified spontaneous emission free-electron laser, Science
  292~(5524) (2001) 2037--2041.

\bibitem{sun2018high}
Y.~Sun, W.~Berg, J.~Byrd, J.~Dooling, D.~Hui, S.~Shin, A.~Zholents, The high
  brightness photo-injector electron beam of the aps linac, in: Future Light
  Source, 2018, p. THP1WD03.

\bibitem{Berg:LEA}
W.~J. Berg, et~al., Development of the linac extension area 450-mev electron
  test beam line at the advanced photon source, in: International Beam
  Instrumentation Conference, 2019, p. mopp048.

\bibitem{BorlandBC2000}
M.~Borland, Design and performance simulations of the bunch compressor for the
  advanced photon source low-energy undulator test line free electron laser,
  Physical Review Special Topics-Accelerators and Beams 4 (2001) 074201.
\newblock \href {https://doi.org/10.1103/PhysRevSTAB.4.074201}
  {\path{doi:10.1103/PhysRevSTAB.4.074201}}.

\bibitem{astra}
K.~Flottmann, S.~Lidia, P.~Piot, Recent improvements to the astra particle
  tracking code 5 (2003) 3500--3502.
\newblock \href {https://doi.org/10.1109/PAC.2003.1289961}
  {\path{doi:10.1109/PAC.2003.1289961}}.

\bibitem{borland_elegant}
M.~Borland, \href{https://www.osti.gov/biblio/761286}{Elegant: A flexible
  sdds-compliant code for accelerator simulation}, Tech. Rep. LS-287, Argonne
  National Laboratory Advanced Photon Source (8 2000).
\newblock \href {https://doi.org/10.2172/761286} {\path{doi:10.2172/761286}}.
\newline\urlprefix\url{https://www.osti.gov/biblio/761286}

\bibitem{england_sextupole}
R.~England, J.~Rosenzweig, G.~Andonian, P.~Musumeci, G.~Travish, R.~Yoder,
  Sextupole correction of the longitudinal transport of relativistic beams in
  dispersionless translating sections, Physical Review Special Topics -
  Accelerators and Beams 8~(1) (2005) 012801.

\bibitem{hall_sextupole}
C.~C. Hall, Study of collective beam effects in energy recovery linac driven
  free electron lasers, Ph.D. thesis, Colorado State University (2016).

\bibitem{bane2016}
K.~Bane, G.~Stupakov, I.~Zagorodnov,
  \href{https://link.aps.org/doi/10.1103/PhysRevAccelBeams.19.084401}{Analytical
  formulas for short bunch wakes in a flat dechirper}, Physical Review
  Accelerators and Beams 19 (2016) 084401.
\newblock \href {https://doi.org/10.1103/PhysRevAccelBeams.19.084401}
  {\path{doi:10.1103/PhysRevAccelBeams.19.084401}}.
\newline\urlprefix\url{https://link.aps.org/doi/10.1103/PhysRevAccelBeams.19.084401}

\bibitem{penco2017passive}
G.~Penco, E.~Allaria, I.~Cudin, S.~Di~Mitri, D.~Gauthier, S.~Spampinati,
  M.~Trov{\'o}, L.~Giannessi, E.~Roussel, S.~Bettoni, et~al., Passive
  linearization of the magnetic bunch compression using self-induced fields,
  Physical Review Letters 119~(18) (2017) 184802.

\bibitem{craievich2010passive}
P.~Craievich, Passive longitudinal phase space linearizer, Physical Review
  Special Topics-Accelerators and Beams 13~(3) (2010) 034401.

\bibitem{fu2015demonstration}
F.~Fu, R.~Wang, P.~Zhu, L.~Zhao, T.~Jiang, C.~Lu, S.~Liu, L.~Shi, L.~Yan,
  H.~Deng, et~al., Demonstration of nonlinear-energy-spread compensation in
  relativistic electron bunches with corrugated structures, Physical Review
  Letters 114~(11) (2015) 114801.

\bibitem{lu2016time}
C.~Lu, F.~Fu, T.~Jiang, S.~Liu, L.~Shi, R.~Wang, L.~Zhao, P.~Zhu, Z.~Zhang,
  D.~Xiang, Time-resolved measurement of quadrupole wakefields in corrugated
  structures, Physical Review Accelerators and Beams 19~(2) (2016) 020706.

\bibitem{bane2012}
K.~Bane, G.~Stupakov, Corrugated pipe as a beam dechirper, Nuclear Instruments
  and Methods in Physics Research Section A: Accelerators, Spectrometers,
  Detectors and Associated Equipment 690 (2012) 106--110.
\newblock \href {https://doi.org/10.1016/j.nima.2012.07.001}
  {\path{doi:10.1016/j.nima.2012.07.001}}.

\bibitem{zhang_slabcorrugated}
Z.~Zhang, K.~Bane, Y.~Ding, Z.~Huang, R.~Iverson, T.~Maxwell, G.~Stupakov,
  L.~Wang, Electron beam energy chirp control with a rectangular corrugated
  structure at the linac coherent light source, Physical Review Special
  Topics-Accelerators and Beams 18~(1) (2015) 010702.

\bibitem{NSudar2020}
N.~Sudar, Y.~Ding, Y.~Nosochkov, K.~Bane, Z.~Zhang, Octupole based current horn
  suppresion in multi-stage bunch compression with emittance growth correction,
  Physical Review Accelerators and Beams 23~(11) (2020) 112802.
\newblock \href {https://doi.org/10.1103/PhysRevAccelBeams.23.112802}
  {\path{doi:10.1103/PhysRevAccelBeams.23.112802}}.

\bibitem{radia}
O.~Chubar, P.~Elleaume, J.~Chavanne, A three-dimensional magnetostatics
  computer code for insertion devices, Journal of synchrotron radiation 5~(3)
  (1998) 481--484.

\bibitem{halbach}
K.~Halbach, Permanent magnet undulators, Le Journal de Physique Colloques
  44~(C1) (1983) C1--211.

\bibitem{fisher:GPTFEL}
A.~Fisher, P.~Musumeci, S.~Van~der Geer, Self-consistent numerical approach to
  track particles in free electron laser interaction with electromagnetic field
  modes, Physical Review Accelerators and Beams 23~(11) (2020) 110702.

\bibitem{genesis}
S.~Reiche, Genesis 1.3: a fully 3d time-dependent fel simulation code, Nuclear
  Instruments and Methods in Physics Research Section A: Accelerators,
  Spectrometers, Detectors and Associated Equipment 429~(1-3) (1999) 243--248.

\bibitem{duris2014high}
J.~Duris, P.~Musumeci, M.~Babzien, M.~Fedurin, K.~Kusche, R.~Li, J.~Moody,
  I.~Pogorelsky, M.~Polyanskiy, J.~Rosenzweig, Y.~Sakai, C.~Swinson,
  E.~Threlkeld, O.~Williams, V.~Yakimenko, High-quality electron beams from a
  helical inverse free-electron laser accelerator, Nature communications 5
  (2014) 4928.
\newblock \href {https://doi.org/10.1038/ncomms5928}
  {\path{doi:10.1038/ncomms5928}}.

\bibitem{tara:napac}
T.~J. Campese, et~al., Strongly tapered helical undulator system for tessa-266,
  in: Proceedings, North-America Particle Accelerator Conference, 2019, p.
  MOZBA3.

\bibitem{Clarke}
J.~Clarke, The Science and Technology of Undulators and Wigglers, Oxford Series
  on Synchrotron Radiation, OUP Oxford, 2004.

\bibitem{varfolomeev1994undulator}
A.~Varfolomeev, A.~Khlebnikov, N.~Osmanov, S.~Tolmachev, Undulator magnetic
  field measurements with the wire deflection method, Nuclear Instruments and
  Methods in Physics Research Section A: Accelerators, Spectrometers, Detectors
  and Associated Equipment 341~(1-3) (1994) 470--472.

\bibitem{osmanov1998further}
N.~Osmanov, S.~Tolmachev, A.~Varfolomeev, Further development of the pulsed
  wire technique for magnetic field and focusing strength measurements in long
  undulators, Nuclear Instruments and Methods in Physics Research Section A:
  Accelerators, Spectrometers, Detectors and Associated Equipment 407~(1-3)
  (1998) 443--447.

\bibitem{d2016ultra}
A.~D'Audney, Ultra-high resolution pulsed-wire magnet measurement system, an,
  Ph.D. thesis, Colorado State University (2016).

\bibitem{warren1988limitations}
R.~Warren, Limitations on the use of the pulsed-wire field measuring technique,
  Nuclear Instruments and Methods in Physics Research Section A: Accelerators,
  Spectrometers, Detectors and Associated Equipment 272~(1-2) (1988) 257--263.

\bibitem{kumar2010analysis}
V.~Kumar, G.~Mishra, Analysis of pulsed wire method for field integral
  measurements in undulators, Pramana 74~(5) (2010) 743--753.

\bibitem{wu2018recent}
J.~Wu, X.~Huang, T.~Raubenheimer, A.~Scheinker, et~al., Recent on-line taper
  optimization on lcls, in: Proceedings, 38th International Free Electron Laser
  Conference, FEL2017, 2018.

\bibitem{halbach1983conceptual}
K.~Halbach, Conceptual design of a permanent quadrupole magnet with adjustable
  strength, Nuclear Instruments and Methods in Physics Research 206~(3) (1983)
  353--354.

\bibitem{marteau2017variable}
F.~Marteau, A.~Ghaith, P.~N'Gotta, C.~Benabderrahmane, M.~Vall{\'e}au,
  C.~Kitegi, A.~Loulergue, J.~V{\'e}t{\'e}ran, M.~Sebdaoui, T.~Andr{\'e},
  et~al., Variable high gradient permanent magnet quadrupole (quapeva), Applied
  Physics Letters 111~(25) (2017) 253503.

\bibitem{CEmma2015}
C.~Emma, K.~Fang, J.~Wu, C.~Pellegrini, {High efficiency, multiterawatt x-ray
  free electron lasers}, Physical Review Accelerators and Beams 19~(2) (2016)
  020705.
\newblock \href {https://doi.org/10.1103/PhysRevAccelBeams.19.020705}
  {\path{doi:10.1103/PhysRevAccelBeams.19.020705}}.

\end{thebibliography}

\end{document}